\documentclass[%
    reprint,
    superscriptaddress,
    amsmath,amssymb,
    prb,
]{revtex4-1}

\usepackage[table]{xcolor}
\usepackage{textcomp}
\usepackage{amsmath}
\usepackage{amssymb}
\usepackage{amsbsy} 
\usepackage{overpic}
\usepackage{graphicx} 
\usepackage{tikz}
\usetikzlibrary{decorations.pathreplacing}
\usepackage{pgf}
\usepackage{footnote}
\usepackage{array}
\usepackage{multirow}
\usepackage{epstopdf}
\usepackage{mathtools}
\usepackage[toc,page]{appendix}

\graphicspath{{Figures/PNG/}{Figures/PDF/}{Figures/EPS/}{Figures/TEX/}{Figures/}}
\usepackage{float} 

\usepackage{xcolor}

\usepackage{hyperref}
\usepackage{cleveref}

\hypersetup{linktocpage}
\hypersetup{ colorlinks = true, citecolor = islamicgreen,  linkcolor=blue } 
\definecolor{islamicgreen}{rgb}{0,.5,0}
\definecolor{webbrown}{rgb}{.6,0,0}
\definecolor{RoyalBlue}{rgb}{0,.5,0}

\newcommand{\ket}[1]{\left|#1\right\rangle}
\newcommand{\bra}[1]{\left\langle #1\right|}

\newcommand{\Tr}{\operatorname{Tr}}

\newcommand{\mean}[1]{\left\langle #1\right\rangle}

\newcommand{\calH}{\mathcal{H}}

\def\ocite#1{{[\onlinecite{#1}]}}
\def\cO{\hat{\mathcal{O}}}
\def\tr{\operatorname{Tr}}
\def\a{\alpha}
\def\bra#1{\langle #1 |}
\def\ket#1{| #1 \rangle}

\def\ni{\noindent}
\def\nn{\nonumber}
\def \m#1{\hat m^{#1}}
\def\a{\alpha}
\def \b{\beta}
\def\g{\gamma}
\def\q#1{\tilde q_{#1}}
\def\p#1{\tilde p_{#1}}

\def\up{\uparrow}

\begin{document}

\title{Scrambling and entanglement spreading in long-range spin chains}

\author{Silvia Pappalardi}
\email{spappala@sissa.it}
\affiliation{SISSA, Via Bonomea 265, I-34135 Trieste, Italy}
\affiliation{Abdus Salam ICTP, Strada Costiera 11, I-34151 Trieste, Italy}

\author{Angelo Russomanno}
\affiliation{NEST, Scuola Normale Superiore \& Istituto Nanoscienze-CNR, I-56126 Pisa, Italy}
\affiliation{Abdus Salam ICTP, Strada Costiera 11, I-34151 Trieste, Italy}

\author{Bojan \v{Z}unkovi\v{c}}
\affiliation{Faculty of Mathematics and Physics, University of Ljubljana, Jadranska 19, SI-1000 Ljubljana, Slovenia}

\author{Fernando Iemini}
\affiliation{Abdus Salam ICTP, Strada Costiera 11, I-34151 Trieste, Italy}
\affiliation{Instituto de Fisica, Universidade Federal Fluminense, 24210-346 Niteroi, Brazil}

\author{Alessandro Silva}
\affiliation{SISSA, Via Bonomea 265, I-34135 Trieste, Italy}

\author{Rosario Fazio}
\affiliation{Abdus Salam ICTP, Strada Costiera 11, I-34151 Trieste, Italy}
\affiliation{NEST, Scuola Normale Superiore \& Istituto Nanoscienze-CNR, I-56126 Pisa, Italy}

\begin{abstract}
We study scrambling in connection to multipartite entanglement dynamics in regular and chaotic long-range spin chains, characterized by a well defined semi-classical limit. For regular dynamics, scrambling and entanglement dynamics are found to be very different: up to the Ehrenfest time, they rise side by side departing only afterward. Entanglement saturates and becomes extensively multipartite, while scrambling, characterized by the dynamic of the square commutator of initially commuting variables, continues its growth up to the recurrence time. Remarkably, 
the exponential growth of the latter emerges not only in the chaotic case but also in the regular one, when the dynamics occurs at a dynamical critical point.
\end{abstract}


\maketitle
\section{Introduction}
Classical systems with long-range interactioruns display many interesting dynamical properties that have been extensively studied since many
decades~\cite{campa2014physics}. In  the quantum domain, instead, long-range systems have been the focus of a great deal of attention 
only lately,  as a result of their experimental simulation with  different platforms~\cite{Albiez_PRL05,blatt2012quantum,Neyenhuis:2016aa,
Jurcevic:2016aa}. These systems allow the controlled study of quantum dynamics in the absence of significant decoherence, a property that 
allows the study of a number of important phenomena as, for example, dynamical phase transitions~\cite{sciolla2011dynamical, Yuzbashyan2006relaxation, zunkovic2016dynamical} 
or the dynamics of correlations~\cite{hauke2013,cevo2016spreading,cong2014persistence,foss2015nearly, luitz2018emergent} in a situation where Lieb-Robinson bounds do not apply~\cite{lieb1972, hasting2004long}.\\


It is well established that understanding the coherent dynamics of a quantum many-body system requires a thorough understanding of the behaviour of its quantum correlations~\cite{amico2008,eisert2010}.  The {\it spreading of quantum correlations} has been the focus of a lot of 
theoretical efforts~\cite{calabresejpa}, starting from the initial important results on the dynamics of entanglement entropy~\cite{calabrese2005}.
Very recently a new way to characterise quantum dynamics of many-body systems has been proposed, based on the concept of \emph{scrambling}. Initially introduced 
as a probe of quantum  chaos~\cite{larkin1969quasiclassical,kitaev2015talk, malda2016syk,kitaev2015talk, maldacena2016bound}, scrambling is generically identified as the delocalisation of quantum information~\cite{Mezei2017entan} in a many-body system.  A measure of scrambling is associated with the growth of the square commutator between two initially commuting observables~\cite{larkin1969quasiclassical,kitaev2015talk}. For quantum chaotic systems \cite{larkin1969quasiclassical,kitaev2015talk,aleiner2016micro,aavis2017chaos}, this quantity is expected to grow exponentially before the Ehrenfest time - defined as the time at which semi-classics breaks and quantum effect become dominant \cite{schubert2012wave}-  otherwise, it grows at most polynomially in time ~\cite{chen2017out,swingle2017slow, kukulian2017weak}. \\

Despite the impressive progress over the last years, several different questions related to scrambling and entanglement propagation still await a more detailed answer. It has been observed that the exponential growth of the square commutator is connected to the chaotic behaviour of an underlying semi-classical limit.  The precise role of semiclassical correlations in determining scrambling dynamics and its various stages are presently under intense study~\cite{rose2017classical, scaffidi2018classical, rammensee2018many}.  Furthermore, in view of the various forms in which quantum correlations manifests in a many-body system, it is important to understand how entanglement is connected to the scrambling of information. A first connection between square commutators and the spreading of quantum entanglement has been made in the context of unitary quantum channels~\cite{hosur2015chaos, ding2016conditional}.  An analysis of different velocities of propagation of information has been performed in~\ocite{Mezei2017entan}, while connections of scrambling to the growth of R{\'e}nyi entropies and multiple-quantum coherence spectra have been investigated in~\ocite{tripartiteandent, tropa2, garttner2019prl}. In long-range systems, scrambling has been studied in connection to correlation bounds \cite{luitz2018emergent, zhou2017measuring} and its time average as a probe of criticality \cite{heyl2018detecting}.
However, an analysis of the dynamics and the relevant time-scales in relation to the different processes involved in the spreading of information is still missing. \\
In this work, we address all these questions by studying multipartite entanglement propagation and scrambling in spin chains with long-range interaction either subject to a quantum quench or a periodic drive.  
There are several reasons behind this choice. Spin chains with long-range interactions possess a well defined semiclassical limit, and thus represent a  natural playground~\cite{swingle2017proposal} to study the role of classical correlations in scrambling. Furthermore, they allow exploring the transition from semiclassical to quantum dominated regimes in the dynamical behaviour. We will consider both the case of integrable and chaotic dynamics. Moreover, 
scrambling is experimentally accessible with long-range quantum simulators, as it has been measured for unitary operators~\cite{gattner2017measuring}.
We will present results for the entanglement dynamics of the quantum Fisher information, the tripartite mutual information and operator scrambling, studied via the square commutator. As we are going to show in the rest of the paper scrambling and entanglement dynamics turn out to be very different. 

\onecolumngrid

\begin{table}[H]
\begin{center}
\begin{tabular}{ l r || r r r }
\hline
\multirow{2}{4.5em}{ $\a=0$ } & & \multirow{2}{9em}{ Quantum quench } & \multirow{2}{9em}{ Quench at DPT} & \multirow{2}{9em}{Periodic kicking}  \\
& & & & \\
\hline \hline
& & & & \\
Scrambling & \quad\quad  $t<t_{\text{Ehr}}$ &$t^2/N^3\quad\quad \quad$& $e^{\lambda t}\quad \quad\quad\quad $& $e^{\lambda t}\quad \quad\quad\quad $\\
& \,$t_{\text{Ehr}}<t<t^*$ & $t^4/N^4\quad\quad \quad$&$t/N\quad \quad\quad \quad$ &const.\quad \quad\quad\quad   \\ 
& & & & \\
Entanglement& \quad\quad  $t<t_{\text{Ehr}}$ &growth\quad\quad\quad \quad & peak\quad \quad\quad\quad \quad & growth \quad \quad\quad\quad \\
& \,$t_{\text{Ehr}}<t<t^*$ & const.\quad\quad \quad\quad & const.\quad\quad\quad \,\,\,  & const.\quad \quad\quad\quad  \\
\end{tabular}
\caption{Scrambling and entanglement dynamics for the different protocols with the infinite range hamiltonian. While for all the dynamics, entanglement grows and saturate at $t_{\text{Ehr}}$, scrambling continues his growth in the regular case. Particularly interesting is that, despite the dynamics being regular, the early-time exponential behaviour emerges when the dynamics occur at the critical point of the dynamical phase transition DPT, see Fig.\ref{fig:otoc_kicked_comparison}. The Ehrenfest time and the recurrence time depend on the dynamics too: $t_{\text{Ehr}}\propto \sqrt N$ for the regular quantum quench, $t_{\text{Ehr}}\propto \log N$ for the quench at DPT and for the periodic kicking, while $t_{\text{rec}}\propto N$ for the quantum quench dynamics and $t_{\text{rec}}\propto \exp(\exp(N))$ for the chaotic period kicking.}
\label{tab:results}
\end{center}
\end{table}

\twocolumngrid

The paper is organized as follows. The next section is devoted to a summary of the results with a direct comparison between multipartite entanglement growth and scrambling. 
 In section \ref{sec:model}, we review the long-range version of the Ising chain and the type of dynamics that are considered across the paper. We recall the semiclassical limit together with the quantum and classical characterization of chaos. In section \ref{sec:quantities}, we briefly review the definitions of the quantities under consideration: the quantum Fisher information, the tripartite mutual information, and the square commutator. In section \ref{sec:methods} we describe the different numerical and the semi-analytical methods used to reproduce the behavior of the square-commutator. 
We first present in section \ref{sec:enta_dyn} the results for the entanglement dynamics and its semiclassical nature for sufficiently long-range interaction. We discuss how this behavior changes when the range of the interaction is decreased. Then, in section \ref{sec:scrambling}, we consider the results for the square-commutator and we argue that the long-time dynamics of the square commutator accounts for the quantum chaoticity of the dynamics. We provide evidence for our claims by discussing an example of an exponential growth of scrambling in the case of a regular quantum dynamics. Section \ref{sec:Conclusions} is devoted to our conclusions.

\section{Main results}
\label{sec:summa_resa}

In this paper, we study how entanglement and operator's scrambling grow and spread in Ising spin chains with two-body power-law decaying interactions, $J_{ij}\propto |i-j|^{-\a}$. We consider the case in which an initial separable state, i.e. $\ket{\psi_0}=\ket{\up\, \up\,\dots \up\,}$, is brought out-of-equilibrium by means of a quantum quench or a periodic drive. Our findings can be summarized as follows:

\begin{enumerate}
\item Entanglement dynamics reflects the semi-classical nature of the system: it is weak, slowly growing and saturating at the Ehrenfest time $t_{\text{Ehr}}$. This is what lies at the heart of the classical ``simulability'' of quantum long-range interacting systems in the context of MPS-TDVP~\cite{haegeman2011time,
haegman2016uni} with small bond-dimension as well as semiclassical methods \cite{polkovnikov2010phase, schachenmayer2015many, wurtz2018clustered}.
\item The square commutator is characterized by two different regimes, a first semiclassical growth up to the $t_{\text{Ehr}}$ (exponential for chaotic dynamics), followed by a fully quantum non-perturbative polynomial growth (saturation for chaotic dynamics), symmetric around $t^*=t_{\text{rec}}/2$ the \emph{recurrence time} $t_{\text{rec}}$. We show that the initial growth encodes the nature of classical orbits and can be exponential also for regular integrable dynamics, provided they have some classical instabilities. Conversely, the second regime accounts for the quantum chaoticity of the dynamics, see table \ref{tab:results}.
\item  The dynamics of the information spreading changes with the range of interaction $\a$. The Hamiltonian with $0\leq\a<1$ is dominated by the classical limit and the structure of entanglement and scrambling is the same as in the infinite range case, see table \ref{tab:results}. For $1\leq\a<2$, the entanglement grows linearly in time and the structure of the asymptotic state is the same as for $\a<1$. For $\a\geq2$ the state displays the typical entanglement dynamics and structure of short-range interacting systems, with negative TMI.
\end{enumerate}
This shows that state's entanglement growth and operator's scrambling are two distinct, apparently disconnected phenomena. Interestingly, this becomes glaringly obvious in the regular regime, rather than in the chaotic one, see Fig.\ref{fig:comparison}. \\ \bigskip \bigskip


\section{The model and out-of-equilibrium protocols}
\label{sec:model}
We consider an Ising chain in transverse field with long-range interactions,
\begin{equation}
\label{eq:H_LMG}
\hat H= - \frac 12 \sum_{i\neq j}^N J_{ij} \hat \sigma_i^z\, \hat \sigma_j^z  -\, h\sum_i^N \hat \sigma^x_i \ ,
\end{equation}
where $\hat \sigma_i^x, \,\hat \sigma_i^z $ are spin operators and  $J_{ij}=J |i-j|^{-\alpha}/N(\alpha)$ and $N(\a)=\sum_{r=1}^{N}1/r^{\alpha}$ is 
the Kac normalization~\cite{kac1963prescription}.  The (solvable) infinite range limit $\a=0$ of Eq.(\ref{eq:H_LMG}) is known as the Lipkin-Meshov-Glick model 
(LMG)~\cite{Lipkin_NucPhys65} and it has been intensively studied out-of-equilibrium~\cite{bapst2012quantum,sciolla2011dynamical,
russomanno2015thermalization, Halimeh2017dynamical, lang2018dyna}. In this case the Hamiltonian conserves the total spin and we restrict our analysis to the sector of the ground-state $S= N/2$.  This has a
semiclassical limit before $t_{\text{Ehr}}$, controlled by  $\hbar_{\text{eff}} =\hbar /N$, where the system can be described classically in terms of only two degrees of freedom $\{Q, P\}$ and 
a classical Hamiltonian~\cite{bapst2012quantum,sciolla2011dynamical,russomanno2015thermalization}. 
For finite $N$, ground states of the Hamiltonian of Eq.(\ref{eq:H_LMG}) can be 
seen as coherent wave-packets with width ${\sigma \sim \sqrt{\hbar_{\text{eff}}}}$ that evolve for short times classically~\cite{sciolla2011dynamical}. The semi-classical dynamics is discussed in details in next section. As far as the dynamics of local observables is concerned, the Hamiltonian of  Eq.(\ref{eq:H_LMG}) is 
found to behave as the infinite range for $\a<1$, as a short-range one for $\a>2$~\cite{zunkovic2016dynamical, cevo2016spreading}.\\

Taking an initial separable state totally polarised along the $z$ axis 
$\ket{\psi_0}=\ket{\up\, \up\,\dots \up\,}$, we probe entanglement dynamics and scrambling with the two following protocols.\\
\paragraph{Quantum quench}
The state $\ket{\psi_0}$ is evolved with the Hamiltonian~(\ref{eq:H_LMG}) with a transverse field $h_f$. For $\a\leq 1$, a special case, important also for the present analysis, is represented by $h_f=h_c=\frac 12$ where a dynamical phase transition (DPT) occurs \cite{zunko1016dyna}, whose origin can be traced back to the corresponding classical dynamics. Away from  the dynamical critical point the Ehrenfest time reads $t^r_{Ehr} \propto \sqrt N$ while at the 
 dynamical critical point  $t^c_{Ehr}\propto\log N$. It was shown in~\ocite{heyl2018detecting} that DPT can be detected with the 
 average value of out-of-time correlators.  \\
\paragraph{Periodic kicking}  In order to address chaotic dynamics in long-range spin system we will also consider the case in which periodic kicks are added to the evolution governed by Eq.(\ref{eq:H_LMG}), with $\alpha=0$. This model, known also as the ``kicked top'' for $h=0$, is a paradigmatic example of the standard quantum chaos~\cite{haake1987classical, haake2013quantum}.  The time-evolution operator over one period reads
\begin{equation} \label{eq:kicking}
  \hat{U}=
  \hat{U}_{\rm k}\exp\left[-i\hat{H}\, \tau\right]\;
  {\rm with}\;\hat{U}_{\rm k}\equiv
  \exp\left[-i\frac {2\, K} N\, \hat S_z^{\,2}\right]\,.
\end{equation}

Depending on the value of the kicking strength $K$, this model is known to exhibit a transition between a regular regime and a chaotic one~\cite{haake1987classical,
haake2013quantum}. When  $K\gg 1$ $\forall h_f$, orbits deviate exponentially in time and $t^c_{Ehr}\propto \log N$.


\subsection{Semiclassical phase-space}
\label{app:class_lim}
\begin{figure}[H]
\centering
\includegraphics[scale  = 0.52]{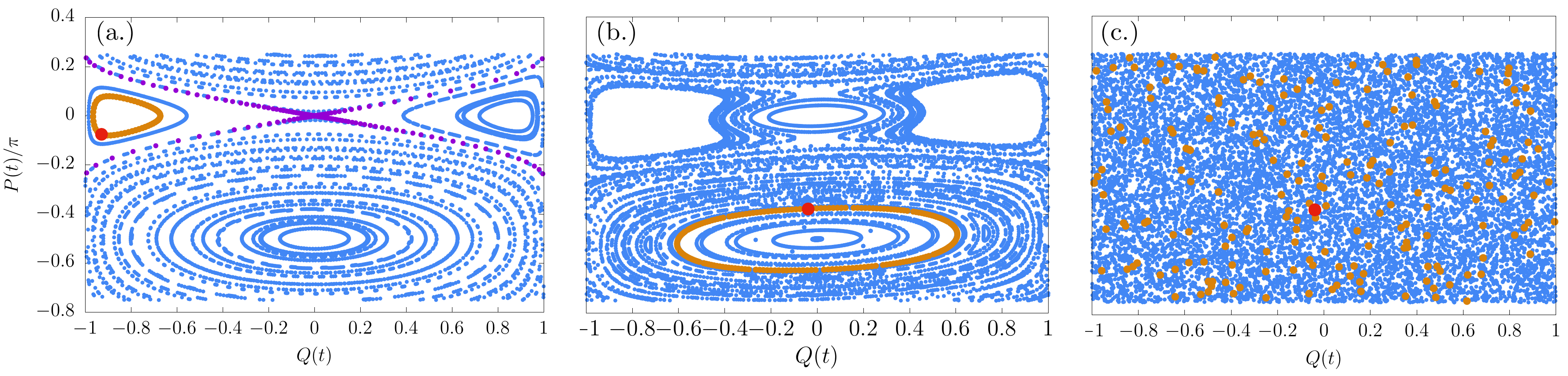}
\caption{Classical phase space (a.) and Poincar{\'e} sections of the classical limit of the model (b., c.). (a.) Phase space for the hamiltonian of Eq.(\ref{eq:h_cla}) for $h_f=\frac 12$. In purple the separatrix between orbits with $\overline Q=0$ and $\overline Q\neq0$, which corresponds at the ground state energy for $h_0=0$. (b.) Regular Poincar{\'e} section for $K =0.2$: we see that the dynamics remains always regular and each trajectory is a closed
curve. (c.) Chaotic  Poincar{\'e} section for $K=20$: the system clearly becomes chaotic and the trajectories tend to cover all the phase space. ($h_f=2,\,\tau=1$). The red dot in the plots represents the initial condition in the classical limit, while the orange points represent its stroboscopic evolution.}
\label{fig:pancarre}
\end{figure}

Let us recall the main features of the semiclassical dynamics. Since the Hamiltonian of Eq.(\ref{eq:H_LMG}) commutes with the total spin $\bold {\hat S}=\sum_i\, \bold{\hat S}_i$, we restrict ourself to the spin sub-sector of the ground state $S= N/2$, where the dimensionality of the Hilbert space is $N+1$. Defining $\bold {\hat m}\equiv \bold {\hat S}/S$, we can re-express the LMG Hamiltonian in terms of its components
\begin{equation}
\label{eq:H_LGM_resca}
    \hat H_{LMG}
    = -\, N\left (
     \frac{J}2 \, \hat m^{z\,\,2} - \, h \, \hat m^x \, \right )
     \ .
\end{equation}
This allows to consider an effective $\hbar_{\text{eff}}= \frac {\hbar}N$ that identifies the semiclassical limit with the large-$N$ mean field one. In what follows we set $\hbar=1$.
In this limit, the system is effectively described by the classical Hamiltonian

\begin{equation}\label{eq:h_cla}
  \calH_0(Q,P)\equiv- \frac{J}{2} Q^2 - {h} \; \sqrt{1-Q^2} \; \cos{(2P)} \ ,
\end{equation}
where the two conjugate variables $Q, \,P$ are given in terms of the expectation values of $\bold{\hat m}$ on a wave packet as $m^z=Q$, $m^x=\sqrt{1-Q^2}\cos(2P)$ and $m^y=\sqrt{1-Q^2}\sin(2P)$ and obey to the classical Hamilton equations,  \cite{bapst2012quantum, sciolla2011dynamical, russomanno2015thermalization}. \\
In the sudden quench case, the ground state at $h_0$ is evolved with the hamiltonian with transverse field $h_f$. The \emph{dynamical phase transition} DPT between a finite and zero order parameter occurs at $h_f=h_c=(h_0+1)/2$. One can define a dynamical order parameter as the average magnetization in time: $\overline {Q}= \lim_{T\to \infty}\int_0^T Q(s) ds$, which is different from zero in the symmetry broken phase.  Indeed, at $h_c$ the phase point associated to the initial ground state energy (which is conserved) lays right on the separatrix of the final Hamiltonian: for $h_f>h_c$ it orbits around the maximum with a $\overline Q=0$, while for $h_f<h_c$ it orbits around one of the two ferromagnetic minima and $\overline Q \neq 0$, see Fig.\ref{fig:pancarre}. \\

When the kicking is added, the total classical Hamiltonian reads
\begin{equation} \label{classical_Ham:eqn}
  \calH(Q,P,t) = \calH_0(Q,P)+\calH_{\rm kick}(Q,P)\sum_n\delta(t-n\tau)\;,
\end{equation}
where $ \calH_{\rm kick}(Q,P) = - \frac{K}{2} Q^2 $: the classical kicking acts every period $\tau$ like a rotation around the $z$ axis with an angle proportional to $m_z$.
In the numerical calculations, we re-express the Hamilton's equation of motion as equation of motions for the spin-components $\bold m$
\begin{align}\label{Eq:diff_magn_class}
    \begin{dcases}
        \dot{m}^x(t) = 2 \, J \, m^y(t)\, m^z(t)\, \\
        \dot{m}^y(t) = 2 h \, m^z(t) - 2\, J \, m^x(t)\, m^z(t)\\
        \dot{ m}^z(t) = -2 h \,  m^y(t) 
    \end{dcases}
    \ .
\end{align}
%

\onecolumngrid

\twocolumngrid


Note that these equations can be obtained also from the expectation value of the Heisenberg equation of motion, setting to zero the second order cumulant. This is justified by the fact that the magnetization components commute in the classical limit: $\left [ \hat m^{\a}, \hat m^{\b}\right ] = \frac i{N/2}\, \epsilon_{\a \b \gamma} \hat m^{\gamma} $. 
In the limit of large but finite $N$, one can consider the semiclassical WKB approximation \cite{sciolla2011dynamical} and explore wave-packet dynamics. In this framework, ground states of $\calH_0$ can be seen as coherent wave packets with width $\sigma=\sqrt{\hbar_{\text{eff}}}= 1/\sqrt N$. 
This semiclassical picture holds until the states behave like well defined wave-packets. It is then natural to define the time for which semi-classics breaks down- the \emph{Ehrenfest time} - as the time for which the initially coherent wave-packet is spread and delocalized. It is well known that this depends on the nature of the classical dynamics \cite{schubert2012wave}
\begin{align}
t_{\text{Ehr}} \sim
\begin{dcases}
\frac 1{\sqrt{\hbar_{\text{eff}}}} = \sqrt N \quad \quad \quad \quad \quad \text{regular }\\
\frac 1{2\lambda} \ln \frac 1{\hbar_{\text{eff}}}= \frac 1{2\lambda} \ln N \quad \quad \text{chaotic/unstable}\\ 
\end{dcases}
\ ,
\end{align}
where $\lambda>0$ is the Largest Lyapunov exponent of the classical dynamics in the chaotic case.


\subsection{Characterization of chaos} 
In the quantum realm, an important signature of chaos is provided by the spectral properties of the evolution operator, in our case by the properties of the Floquet spectrum. The distribution of the Floquet level spacings $\delta_{\a}\equiv\mu_{\a+1}-\mu_{\a}$ (the $\mu_{\a}$ are in increasing order), normalized by the average density
of states, gives information on the integrability and ergodicity
properties of the system  \cite{haake2013quantum, berry2983semicla, bohigas1984chaos, kos2018quantum}: if the distribution is Poisson, then the system is integrable; if it is Wigner-Dyson,
then the system is ergodic. 
\begin{figure}[t]
\centering
\fontsize{13}{10}\selectfont
\includegraphics[width = 85 mm ]{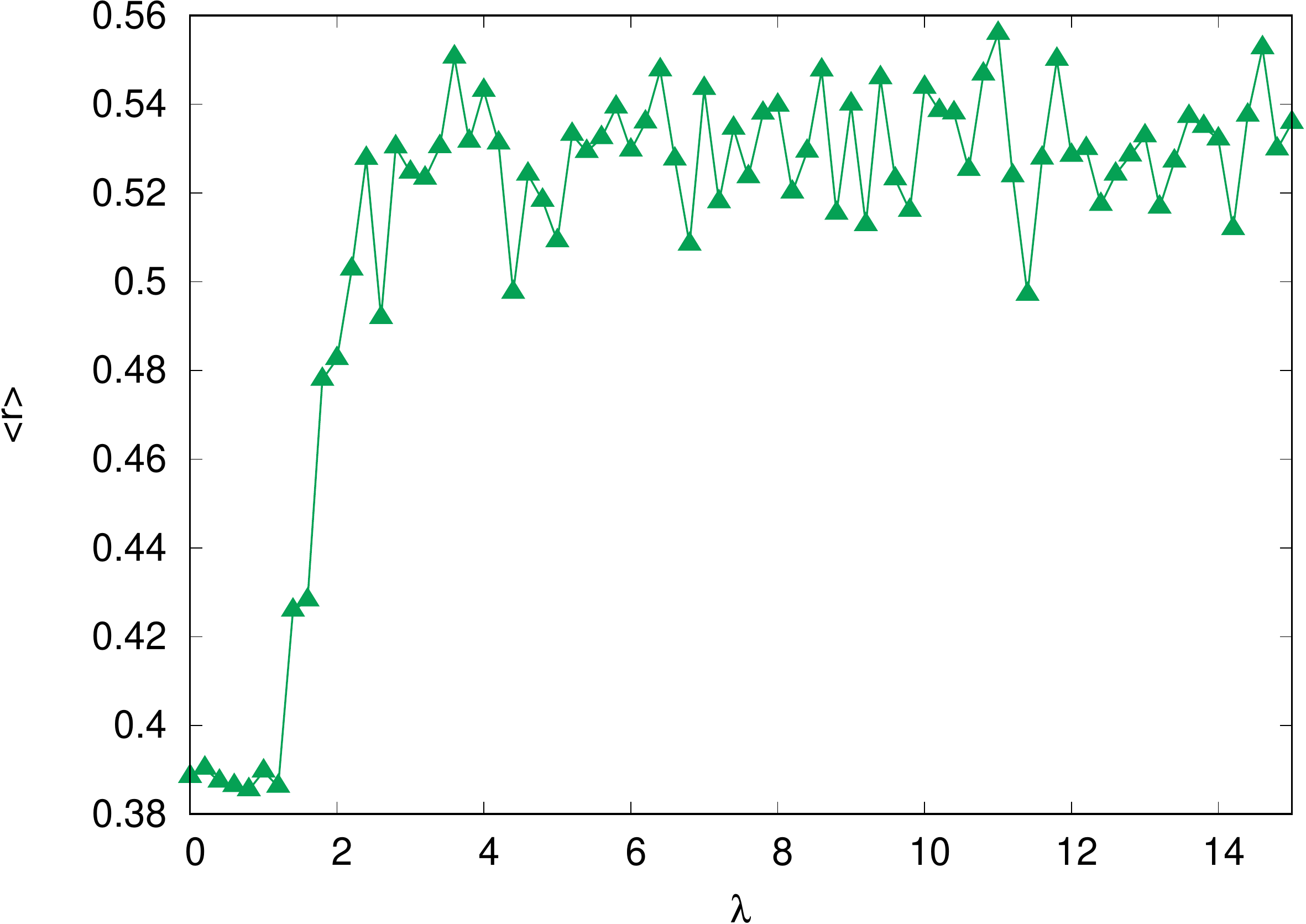}
\caption{Regular-chaotic transition witnessed by the average level spacing ratio. ($N=1000, \,\tau=1,\, h=2$). Throughout the paper, we always chose $K=20$, which clearly displays chaotic dynamics. }
\label{fig:ratio}
\end{figure}
In order to probe the integrability/ergodicity properties through the level spacing distribution, we consider the so-called level spacing ratio
\begin{equation}
  0\leq r_\alpha \equiv \frac{\min\left\{\delta_\alpha,\delta_{\alpha+1}\right\}}{\max\left\{\delta_\alpha,\delta_{\alpha+1}\right\}}\leq 1\,.
\end{equation}
The different level spacing distributions are characterized by a different value of the average $r\equiv\mean{r_\alpha}$ over the distribution. From
the results of Ref.~\ocite{oganesyan2007localization}, we expect $r=0.386$ if the system behaves integrably and the distribution is Poisson; on the other side, if the distribution is Wigner-Dyson and the system behaves ergodically, then $r=0.5295$. In our case, the Floquet levels fall in two symmetry classes, according with the corresponding Floquet state being an eigenstate of eigenvalue $+1$ or $−1$ of the operator $e^{i \pi \hat S_x}$ under which the Hamiltonian is symmetric \cite{haake1987classical}. Therfore we need to evaluate the level spacing distribution and the corresponding  $r$ only over Floquet states in one of the symmetry sectors of the Hamiltonian.
The level spacing ratio for this model is reported in Fig.\ref{fig:ratio} as a function of the kicking strength $K$: it shows a transition from a regular to a chaotic regime.

The relationship between classical chaos and the properties of the many-body quantum dynamics has been widely studied in the past, giving rise to a plethora of signatures of chaos in the quantum domain \cite{haake2013quantum, berry2983semicla}. 
Classically, a system is ergodic if all the trajectories uniformly explore the accessible part of the phase space. In case of few degrees of freedom, a qualitative measure of this phenomenon is the Poincar{\'e} section:
some initial values are evolved under the stroboscopic
dynamics reporting on a $P, Q $ plot the sequence of their
positions. If the initial condition lies in a regular region of
the phase space, our points will be over a one-dimensional
manifold. If instead, the initial condition is in a chaotic region of
the phase space, our points will fill a two-dimensional portion
of phase space. The model under analysis satisfies this conditions in the semi-classical limit, see Fig.\ref{fig:pancarre}.\\

\section{Characterization of entanglement and scrambling}
\label{sec:quantities}
Let us now introduce the quantities that we will use to characterize entanglement and scrambling. 
As far as the entanglement is concerned, we will focus on the multipartite case (bipartite entanglement was already studied  in~\ocite{Schachenmayer2013entanglement,Buyskikh2016entanglement}).  The characterisation of multipartite entanglement is more delicate than that of bipartite entanglement since there exist a zoo of possible measures and witnesses. We will focus here on the quantum Fisher information (QFI) $F_Q(t)$ and on the tripartite mutual information (TMI) $I_3(t)$~\cite{hosur2015chaos}, which accounts for the information delocalization. Scrambling is instead studied via the square commutator $c(t)$.\\

The \emph{quantum Fisher information} is a witness of multipartite entanglement which has been shown 
to obey scaling at the equilibrium transition point ~\cite{hauke2016measuring} and is connected to the diagonal ensemble in the 
non-equilibrium case ~\cite{pappalardi2017multipartite}.
The QFI gives a bound on the size of the biggest entangled block. For example, given a system 
of $N$ spins, if the QFI density $f_Q\equiv F_Q/N>k$, then there are at least $k+1$ entangled spins~\cite{hyllus2012multi, toth2012multi}.
For pure states, the QFI is given by an optimization over a generic linear combination of local spin operators of $F_Q(\cO, t)=\, 4\, \langle \Delta \cO^2 \rangle_t$. 
Here, we consider collective spin operators $\cO=\bf{ \hat S}=\frac 12 \sum_i\bf{\hat \sigma}_i$ and we maximise over the three directions. \\

The \emph{tripartite mutual information} is defined as ${ I_3(A:B:C)= I(A:B)+I(A:C)-I(A:BC) }$ where $A,\,B, \, C, \, D$ are four partitions and the quantity $I(A:B)$ is the mutual information between $A,\,B$. This takes into account information about A that is non-locally stored in $C$ and $D$ such that local measurements of $B$ and $C$ alone are not able to re-construct $A$. Usually $I_3<0$ is associated with the delocalisation of quantum information in the context of unitary quantum channels \cite{hosur2015chaos}. In this case, more appropriately,  we study the delocalisation of the initial state information under the dynamics, which is a complementary measure of entanglement.\\
  
Finally, in order to characterise the dynamics of scrambling, we will focus on the \emph{square commutator} $c(t)=-\langle [\hat B(t), \hat A]^2\rangle$. This object measures the non-commutativity induced by the dynamics between two initially commuting operators $\hat A$ and $\hat B$. It was introduced by Larkin and Ovchinnikov in \ocite{larkin1969quasiclassical}, to describe semiclassically the exponential sensitivity to initial conditions and the associated Lyapunov exponent.
By taking collective spin operators ~\cite{kukulian2017weak} $\hat A=\hat B =\hat m_z= \hat S_z/S$, the square commutator has a natural classical limit for $\hbar_{\text{eff}}\to 0$ ~\cite{cotler2017out}
\begin{eqnarray}
\label{eq:otoc} 
c(t) =
    -\,\langle \, \left [ {\hat m^z(t)}, {\hat m^z} \right ]^2\, \rangle
      \to  \hbar_{\text{eff}}^2  \overline{\{Q(t), Q(0)\}^2} \ ,
\end{eqnarray}
where $Q(t)=\langle \hat m_z(t)\rangle$ on a coherent wave packet, $\{\cdot\}$ are the Poisson brackets of the corresponding classical trajectory and the average $\overline {(\cdot)}$ is performed over an initial phase-space distribution.   \\

\section{Methods}
\label{sec:methods}

\begin{figure}[t]
\centering
\includegraphics[scale  = 1]{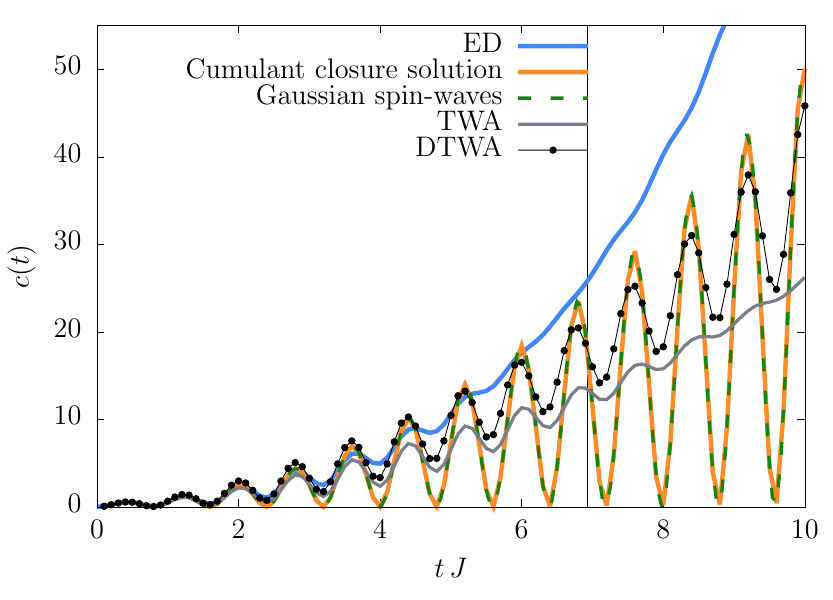}
\caption{Comparison between the different semi-classical approximations of the square commutator. Quantum quench dynamics with $h_f=2$ and $N=50$. All the approximated methods are found to reproduce the dynamics of $c(t)$ up to times $t^r_{Ehr}$, indicated with a thin line in the plot. }
\label{fig:compa_exact_approxi}
\end{figure}

The results presented in this work were obtained with a series of numerical techniques and two semi-analytical approximations. \\
The numerical methods are a combination of exact diagonalization (ED) and well-established semi-classical approximations which are based on Wigner phase-space representations: the truncated Wigner approximation (TWA)~\cite{Blakie2008dynamics} on the continuum phase-space and the discrete truncated Wigner approximation (DTWA) \cite{wootters1987discrete,schachenmayer2015many} of the finite dimensional phase space. 
To this end, we generalised the corresponding expression for the square commutator to the discrete phase space representation, see Eq.(\ref{eq:otoc_DTWA}) and the supplementary material for the details used in our calculations. All these approaches neglect terms of the order of $\mathcal O(1/N)$ and give the same results up to the Ehrenfest time. DTWA, in particular, is able to reproduce also entanglement long-time dynamics.  
Furthermore, we also adopted the matrix product state time-dependent variational principle (MPS-TDVP) \cite{haegman2016uni, haegeman2011time}, for the dynamics of long-range hamiltonians with $\a\neq 0$. \\
We combine these approaches with two semi-analytical methods in order to predict the behavior of $c(t)$ up to $t_{\text{Ehr}}$.
 The first method is a \emph{equation of motion closure at fifth order}: it consists in deriving a hierarchy of differential equations for the square commutator and in closing it by setting the fifth order cumulant to zero. This allows to decouple the higher order commutator and to close the system of equations. By setting the appropriate initial conditions, one can integrate numerically the equations and get the approximated $c(t)$. The second method is a \emph{time-dependent Holstein Primakoff} and it consists in including quantum fluctuations on top of the classical result and to keep it only at the Gaussian level. These approaches turn out to be equivalent and to correctly reproduce $c(t)$ before $t_{\text{Ehr}}$ as in Fig.\ref{fig:compa_exact_approxi}. The following two paragraphs are devoted to a description of these approximations.
%
\subparagraph{Equation of motion closure at fifth order}
The cumulant closure at order $n$ is a general method that consists in closing a set of differential equations, by setting to zero all cumulants of the order $\geq n $. A very easy example is the cumulant closure at second order, which allows computing the classical equation of motion for the magnetization of Eq.(\ref{Eq:diff_magn_class}).
We are interested in the dynamics of the square commutator and we wish to find a set of differential equations that gives its evolution. \\

One first defines a symmetric $(n+m+2)$-string commutator
\begin{align} \label{eq:string_commutator}
c_{\a_1, \dots \a_n , \b_1 \dots \b_m}(t)  = & 
-\frac 12 \langle\, \left[\m{\a_1}(t) \, \m{\a_2}(t) \dots \m{\a_n}(t), \m z\right] \times\,\nn \\
& \quad \quad \, \left[\m{\b_1}(t) \, \m{\b_2}(t) \dots \m{\b_m}(t), \m z\right ] \, \rangle \nn \\
& -\frac 12 \langle\, \left[\m{\b_1}(t) \, \m{\b_2}(t) \dots \m{\b_m}(t), \m z\right ]\times \nn \\
& \quad \quad \, \left[\m{\a_1}(t) \, \m{\a_2}(t) \dots \m{\a_n}(t), \m z\right]\, \rangle  \ ,
\end{align}
where there are $n+m$ time-dependent operators and two time-independent ones. Within this notation, the square commutator of Eq.(1) reads $c(t)=c_{z,z}(t)$. The dynamics will generate an infinite number of coupled equations of motion. We close this hierarchy of differential equations by setting the fifth order cumulant to zero: $\langle A B C D E \rangle_c=0$. If one assumes that the magnetization is classical (second order cumulant set to zero), the $(3+2)$-string commutator decouples in 
\begin{equation}
\label{eq:wowowo}
c_{\a, \b\g}(t) + c_{\a, \g\b}(t) = 2 \left[ \,\,
m^{\b}(t)\,  c_{\a,\g}(t) + m^{\g}(t)\,  c_{\a,\b}(t)
\,\,\right ] \ ,
\end{equation}
for $\a,\b,\g\in\{x, y, z\}$. This allows to close the hierarchy of differential equations, which are coupled to the classical magnetization dynamics of Eq.(\ref{Eq:diff_magn_class}) as
\begin{align}
\begin{dcases}
    \dot c_{z,z} = -4\, h\, c_{z,y}   \\
    \dot c_{z,y} = -2\, h\, c_{y,y} +2\, h\, c_{z,z} -2\, J \left [ c_{z, z}\, m^z + c_{z, x}\, m^x \right ]\\
    \dot c_{y,y} = 4\, h\, c_{y, z} - 4\, J  \left [ c_{x, y}\, m^z + c_{x, y} \, m^z \right ]\\
    \dot c_{x,y} = 2\, h\, c_{x, z} -2\, J \left [ c_{x, x}\, m^z + c_{x, z}\, m^x\right ] \\ \quad \quad \,
    +2\, J \left [ c_{y, y}\, m^z + c_{y, z}\, m^y\right ] \\
    \dot c_{x, z} = - 2\, h\, c_{x, y} + 2\, J \left [ c_{z, z}\, m^y + c_{z, y}\, m^z\right ] \\
    \dot c_{x, x} = 2\, J \left [ c_{x, z}\, m^y + c_{x, y}\, m^z\right ] \ . \\
\end{dcases}
\end{align}
These equations are integrated numerically via a fourth order Runge-Kutta method. The information about the initial state and the dimension of the system is encoded in the initial conditions.  The $c_{z,z}(t)$ that we obtain with this cumulant closure turns out to reproduce the limit $N\to\infty$ of all these semiclassical approximations. It well reproduces the exact $c(t)$, up to a time $t^r_{Ehr}$ (see Fig.\ref{fig:compa_exact_approxi}).

\subparagraph{Time-dependent Holstein Primakoff}
In a spin wave expansion, quantum fluctuations are treated as small fluctuations on top of the classical solution~\cite{lerose2017chaotic}. 
One first produces a time-dependent rotation of the reference frame $\mathcal R=(\hat X(t), \hat Y(t), \hat Z(t))$, in such a way that the $\hat Z(t)$ axis follows the motion of the classical collective spin $\langle{\bf \hat S}(t) \rangle$. Then an Holstein-Primakoff transformation is performed and the quantum fluctuations are kept at the gaussian order. In this rotating frame the collective spin operators are the zero-mode components in the Fourier transform: $\tilde \sigma^0_a$, with $a\in \mathcal R$. Our approximation consists in taking the operator on the $Z$-axis not varying in time: $\tilde \sigma^0_Z(0)\sim\tilde \sigma^0_Z(t)+ \mathcal O((N/2)^{-1})$. This allows to compute commutators in the rotating frame, hence to get an approximated solution for $c(t)$. \\
The main steps to solve the dynamics are the following \cite{lerose2017chaotic}:
\begin{itemize}
\item perform a time dependent unitary rotation with $V(\, \theta(t), \phi(t)\, ) = e^{-i \frac{\phi(t)}2 \sum_i  \hat {\sigma}_i^z}e^{-i \frac{\theta(t)}2 \sum_i  \hat {\sigma}_i^y}$; the angles are chosen such that $\langle \hat S_X\rangle = \langle \hat S_Y\rangle=0$. In the new frame the operators evolve with $\tilde H = V\, H\, V^{\dagger} +i V \dot{V}^{\dagger}$;
\item perform an Holstein-Primakoff transformation on the operators in $\mathcal R$ in terms of the conjugate variables $(\tilde q_0, \, \tilde p_0)$;
\item keep only Gaussian terms, which is equivalent to neglect all $\mathcal O((N/2)^{-3/2})$ terms in the equations.
\end{itemize}
With such a choice one remains with the following hamiltonian
\begin{equation}
\label{eq:rota_ham}
\frac {\tilde H}N   
= h_{\text{class}}(t) + \frac 1{\sqrt {Ns}}\, h_{\text{lin}}(t) + \frac 1{Ns}\, h_{\text{quad}}(t) +\mathcal O({(Ns)}^{-3/2}) \ ,
\end{equation}

Then, by setting $\langle \hat S_X\rangle = \langle \hat S_Y\rangle=0$, one gets the \emph{equation of motion for the rotating frame}, see \ocite{das2006infinite}
\begin{align}
\label{eq:motion_angles0}
\begin{dcases}
\dot \theta  =  2 J \sin\theta \cos\phi \sin\phi  \\
\dot \phi = -2h  + 2 J   \cos\theta \cos^2\phi   \ .
    \end{dcases}
\end{align}
In the same way one can obtain the Heisenberg equation of motion for for $\tilde q_0,\,\tilde p_0$. Further defining the zero-mode fluctuations as
\begin{subequations}
\label{eq:quantum_fluctua}
\begin{align}
\Delta^{qq}_0(t)  & \equiv  \langle \, \q{0}(t) \, \q{0}(t)\, \rangle \\
\Delta^{pp}_0(t)  & \equiv \langle \, \p{0}(t)\,  \p{0}(t)\, \rangle \\
\Delta^{qp}_0(t)  & \equiv  \frac 12 \langle \, \q{0}(t) \, \p{0}(t) + \, \p{0}(t)\,  \q{0}(t)\rangle \ .
\end{align}
\end{subequations}
and combining them with the equations for $\tilde q_0,\,\tilde p_0$, one gets the \emph{equations of motion for the zero-mode fluctuations}
\begin{align}
\label{eq:motion_feedback_0}
\begin{dcases}
\dot{\Delta}_0^{qq} = 4J \cos\theta\sin\phi\cos\phi\, \Delta_0^{qq} + 4 J  \left(\cos^2\phi-\sin^2\phi \right)\,\Delta_0^{pq} \\
\dot{\Delta}_0^{pp} = -4J \cos\theta\sin\phi\cos\phi\, \Delta_0^{pp} - 4 J  \cos^2\phi\sin^2\theta\,\Delta_0^{pq}  \\
\dot{\Delta}_0^{pq} = -2J \cos^2\phi\sin^2\theta\,\Delta_0^{qq}  + 2J  \left(\cos^2\phi-\sin^2\phi \right)\,\Delta_0^{pp} \\
\end{dcases}
\end{align}
They are a set of linear time-dependent differential equations, which can be solved numerically with the appropriate initial conditions. They are exactly the quantities that appear in the computation of the square commutator.
In order to compute it perform first a rotation $\tilde \sigma_0^{\a}(t)$ to $\tilde \sigma_0^{a}(t)$ with $V(\theta(t), \phi(t))$, then compute commutators like $[\tilde \sigma_0^{a}(t), \tilde \sigma_0^{Z}]$, noticing that $\tilde \sigma_0^{Z}(0)= \tilde \sigma_0^{Z}(t) + \mathcal O((Ns)^{-1})$, hence at this order they are equal-time commutators that give rise to the zero mode fluctuations of Eq.(\ref{eq:quantum_fluctua}). For example our square commutator of Eq.(3) reads as
\begin{equation}
c(t) = \sin\phi^2\, \Delta^{pp}_0 + \cos^2\theta\cos^2\phi\, \Delta_0^{qq} -2 \cos\theta\sin\phi\cos\phi \, \Delta_0^{pq} \ ,
\end{equation}
which can be obtained numerically from the integration of Eq.(\ref{eq:motion_feedback_0}) and gives exactly the same result of the previous approximation, see Fig.\ref{fig:compa_exact_approxi}. This is correct until the spin-wave density remains small, which, for finite $N$, occurs before $t_{{Ehr}}$. \\
Notice that this method could be in principle extended to long range systems with $\alpha\neq 0$ and other variations of fully connected models \cite{lerose2017chaotic}. In addition one could in principle go beyond the gaussian approximation by keeping the interaction between spin waves.

%
\section{Results} 

\begin{figure}[t]
\centering
\includegraphics[scale  = 0.75]{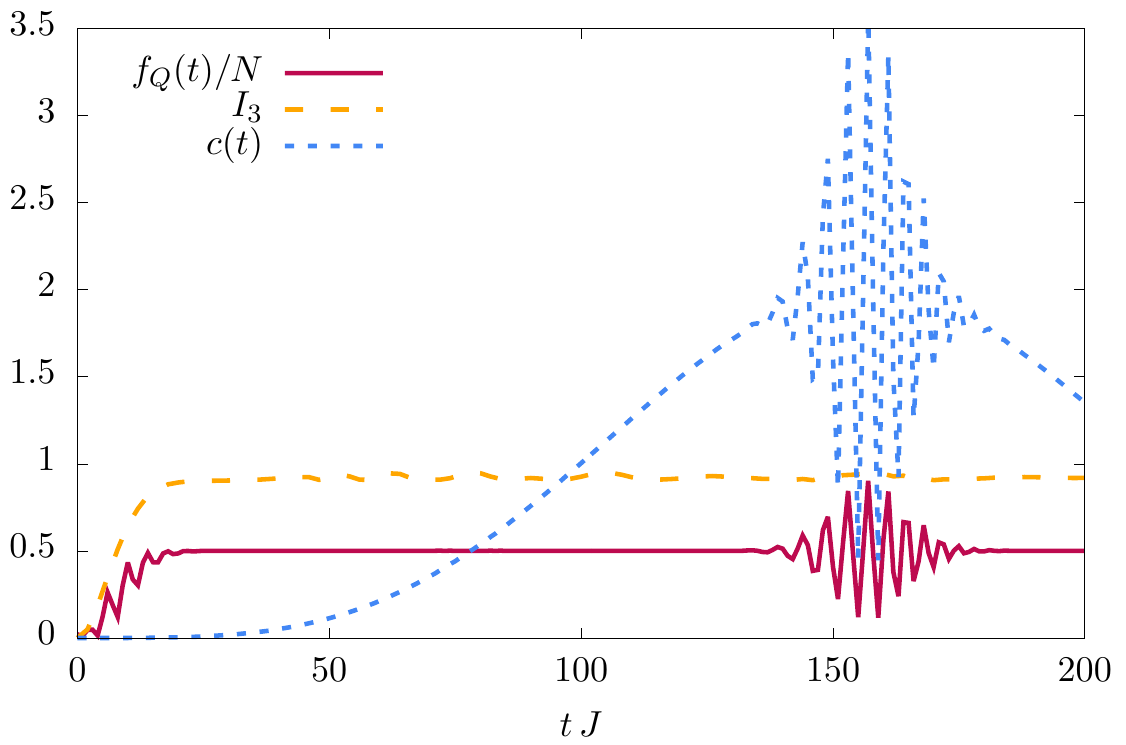}
\caption{Quantum information's dynamics for the regular dynamics. The entanglement quantities, QFI and TMI (red and yellow), saturate at $t_{\text{Ehr}}$, while the square commutator of the longitudinal magnetization operator (blue) goes beyond semi-classics and keeps growing up to $t^*$. Exact diagonalization results for $N=100$, $h_f=2$, TMI with $n_A=1$ $n_B=10$ and $n_C=20$. }
\label{fig:comparison}
\end{figure}
As we hinted out at the beginning of the paper, entanglement and scrambling are two different phenomena, characterized by different time scales, see Fig.\ref{fig:comparison}. 
Let us now finally describe in details the results obtained for the dynamics of entanglement and scrambling using the methods described before.

\subsection{Entanglement dynamics}
\label{sec:enta_dyn}

In the infinite range model, entanglement dynamics and information delocalisation reflect the semiclassical nature of the system under analysis. We start discussing the dynamics governed by Eq.(\ref{eq:H_LMG}) after a quantum quench and describe afterward the case of the periodic kicking protocol. \\

\begin{figure}[t]
\centering
\includegraphics[scale  = 1]{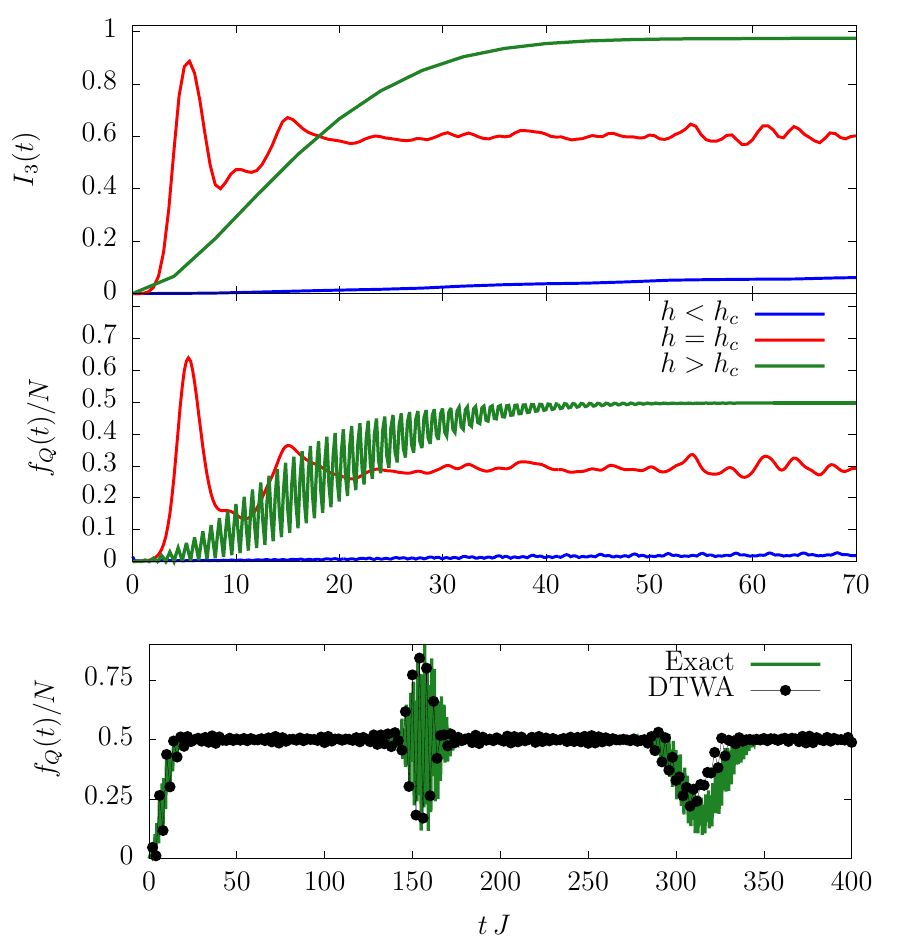}
\caption{QFI and TMI (top and middle) during the evolution after a quantum quench, performed below, at and above the DPT. Entanglement grows in time up to $t^r_{Ehr}$ for quenches above and below the dynamical phase transition (green and blue) and at $t^c_{Ehr}$ at the critical point. Long-time dynamics (bottom) of the QFI, compared with the semiclassical approximation. The DTWA is able to reproduce the dynamics up to $t^r_{\text{rec}}$ and beyond. (Top and middle) $N=450$, $h_f=0.2$, $h_f=0.5$ and $h_f=2$; for the $I_3$: $n_A=1, \, n_B=50,\, n_C=200$. (Bottom)  $N=100$ and $h_f=2$,  DTWA obtained with $5\cdot 10^3$ samplings. }
\label{fig:1}
\end{figure}

\begin{figure}[t]
\centering
\includegraphics[scale  = 1]{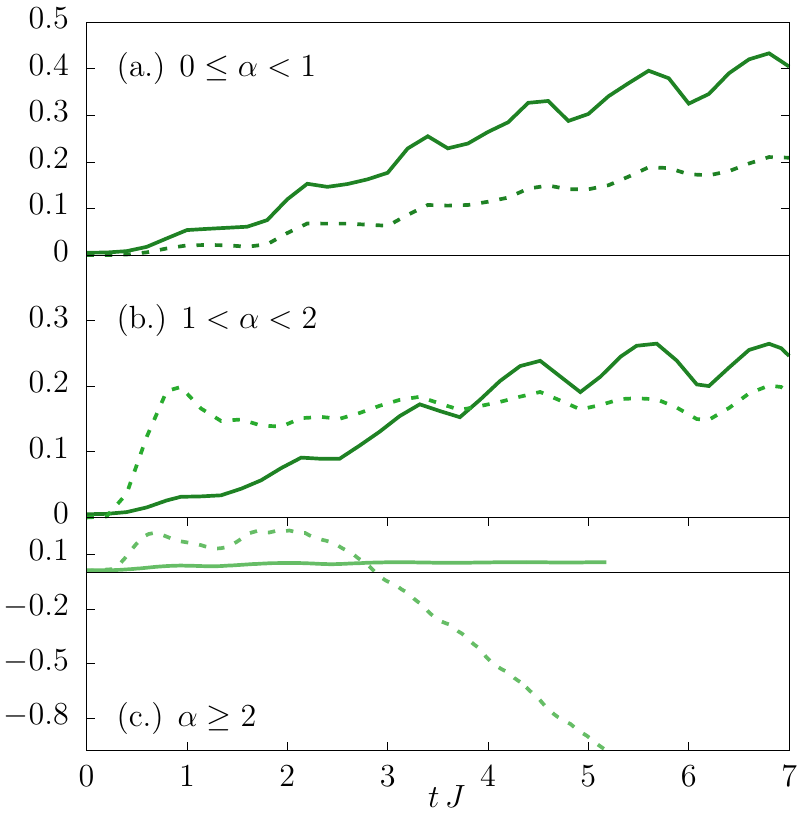}
\caption{ 
We plot the minimal TMI (dotted lines) and the $f_Q(t)/N$ (full lines) as a function of time, varying the range of interaction $\a$.  We obtained the minimal TMI by calculating the tripartite mutual information for all possible partitions A,B,C,D of the system and then taking the minimum. For $0\leq \a<1$ (a.) the dynamics is the same as the LMG model. For $1\leq \a<2$ (b.) we observe linear growth of the QFI with time, whereas the minimum of the tripartite mutual information remains bounded with time. When $\a>2$ (c.), the QFI does not scale with the system size and remains bounded, whereas the minimum of the TMI decreases linearly with time and becomes negative for longer times. We present the detailed scaling with the system size in the supplementary material. The parameters of the evolution are: $\a=0.5$, $\a=1.5$, $\a=2.5$, $h_f=0.75$. Data obtained with TDVP for system sizes $N=200$ and bond dimension $D=256$ (or $N=100$, $D=512$ for $\alpha=2.5$).
}
\label{Fig:bojan}
\end{figure}

Let us focus first on the LMG model at $\a=0$.
Both $f_Q(t)$ and $I_3(t)$ have the same dynamics; growth followed by saturation at $t_{\text{Ehr}}$, as dictated by the semiclassical dynamics of the model, see Fig.\ref{fig:1} (top and middle panels). 
The stationary state displays global
entanglement of genuine multipartite nature $\overline{f_Q}=\phi_Q\, N$,  where $\phi_Q\leq 1/2$  is a function of the transverse field. The value of the phase $\phi_Q$ along the $z$ direction can be computed analytically in terms of elliptic integrals. Following Ref. \ocite{das2006infinite}, with a combination of the classical equation of motion and energy conservation, defining $k=J/2h\geq1$, one gets

\begin{equation}
\phi_Q^z = \frac 1{k^2} \left [\,
    (k^2-1) + \frac{E(\theta_k, k)}{F(\theta_k, k)} 
    - \left (\, \frac \pi {2 F(\theta_k, k)} \right ) ^2
    \, \right ] \ ,
\end{equation}
where $F(\phi, k),\, E(\phi, k)$ are the elliptic integrals of first and second kind of amplitude $\phi$, modulus $k$ and $\theta_k= \arcsin (1/k)$ is the inversion point of the classical trajectory $Q(t)$. 
The maximum asymptotic entanglement witnessed by the QFI is $\overline {f_Q}= \frac {N}2$, which occurs when the system from a product state is quenched to the maximally paramagnetic phase and corresponds to the biggest fluctuations of the collective spin operators.\\
The TMI gives complementary information:  being positive, $\overline {I_3}>0$ it shows that the information of the initial state is not
delocalised across the system. Interestingly, by increasing $\a$ the TMI becomes negative
$I_3<0$, see Fig.\ref{Fig:bojan}. \\

Let us spend a few words for the quench to the DPT, which occurs at $h_c=1/2$, see Sec.\ref{sec:model}. In this case, the entanglement dynamics is qualitatively different. QFI and TMI at short times peak at $t^c_{Ehr}$. After a transient they reach their stationary value, which keeps oscillating without recurrences, see Fig.\ref{fig:1}. This behavior is tightly linked to the existence itself of the DPT, that corresponds to a classical separatrix in phase space: the effective classical trajectory takes time of the order of $\log N$ to depart from its initial value. After that, the classical picture is lost and the state is spread over the basis giving a constant entanglement, see Fig.\ref{fig:enta_dpt}. 

The entanglement dynamics is reproduced, up to very long times, by a semiclassical approach. We studied this regime using DTWA, spin-wave theory and cumulant closure methods, see Sec.\ref{sec:methods}.  All these approaches neglect terms of the order of $\mathcal O(1/N)$ and give the same results up to the Ehrenfest time. 
The accuracy of all the semiclassical analysis is justified by the entanglement structure itself. In fact, this is what lies at the heart of the classical ``simulability'' of quantum long-range interacting systems in the context of MPS-TDVP~\cite{haegeman2011time,
haegman2016uni} and with semiclassical methods \cite{polkovnikov2010phase, schachenmayer2015many, wurtz2018clustered}. DTWA, in particular, is able to reproduce
also the long-time dynamics even beyond the recurrence time $t^r_{\text{rec}}\propto N$ as shown in  Fig.\ref{fig:1} (bottom panel). This is due to the fact that method averages over an extensive number of trajectories, hence mimicking a discreteness of the spectrum, responsible for the recurrences~\cite{schachenmayer2015many}.\\
The same asymptotic structure and dynamics is found for all mean-field like systems 
$0\leq\a<1$: the QFI grows linearly in time up to a value $\sim N$, and the TMI increases logarithmically in time up to a constant value. For $1\leq\a<2$, the QFI and the TMI grow linearly in time and the entanglement structure of the asymptotic state is the same as for $\a<1$. Decreasing the range of interaction the situation changes drastically: for $\a\geq2$ the state displays the typical dynamics and structure of short-range interacting systems 
$\overline{f_Q}\sim \text{const}$; interestingly $\overline{I_3}<0$ signaling that the information about the initial condition is spread throughout the degrees 
of freedom of the state (see Fig.\ref{Fig:bojan}). The results are obtained with TDVP, see the supplementary material for a discussion of the convergence of the method. \\
\begin{figure}[t]
\centering
\includegraphics[scale  = 1.06]{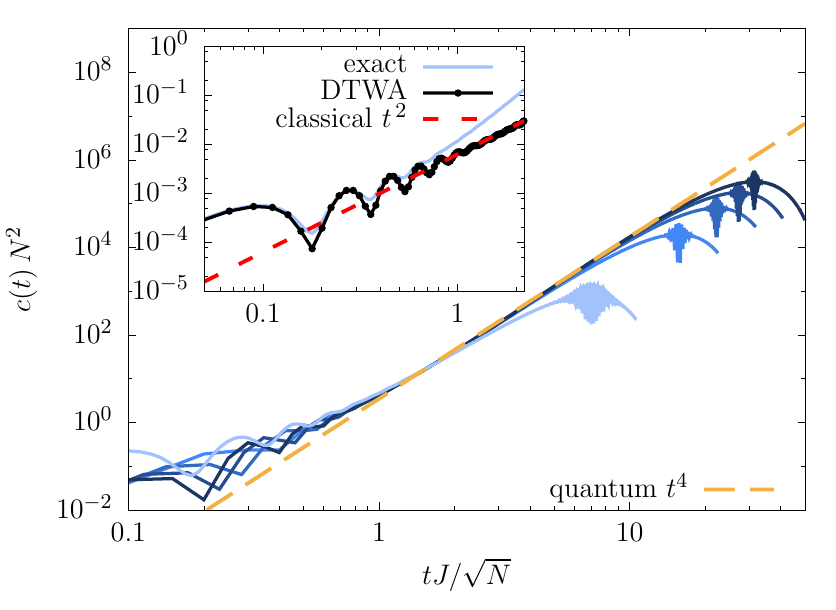}
\caption{The two-times regime of the square commutator in the LMG model after a quantum quench with $h_f=2$. In the main plot the quantum regime of $c(t)$ for different $N$: this regime starts at $\sqrt N$ and then $c(t)$ grows polynomially as $t^4$.  ED results in blue for $N=20, 100, 200,300, 400$ (increasing color darkness), dashed yellow line for the polynomial fit. In the insert we show the semiclassical regime, comparing the exact  $c(t)$ with DTWA, which predicts the $\sim t^2$ power-law growth, dashed in the plot. Here $N=20$ and DTWA obtained with $5\,\cdot 10^3$ samplings.}
\label{fig:otoc_class}
\end{figure}
%
Finally, we conclude the analysis of the multipartite entanglement by considering  the  kicked case in the regime when the dynamics is chaotic. This system heats up to a state where all local observables on any Floquet state correspond to the infinite temperature values~\cite{haake1987classical}.  All quantities characterizing entanglement quantities saturate to an asymptotic value at the Ehrenfest time $t^c_{Ehr}$, for every initial state and field $h$, see Fig.\ref{fig:enta_chaos} of the supplementary. The value of the QFI, being a sum of local observables, is compatible with the values of the infinite temperature state: $\,    \overline{f_Q} = 1 + \frac N 3 + \mathcal O (1/N)$. On the other side, the entanglement entropy saturates to the value expected for a random state, which was derived by Page in \ocite{page1993random}
$ S_{Page}= \log m -\frac m{2 n} + \mathcal O(1/mn) \ ,$ with $m,\,n$ the dimensions of the Hilbert space of the two subsystems and $m\leq n$. In this case, for a partition of size $L$ the dimensions are $m=L+1$, $n=N-L+1$ and $S_{Page}=\log (L+1)+\mathcal O(1/N)$. This reflects on the TMI and we find 
\begin{align}
&\overline{I_3} = \log (\tilde n)\quad \text{with}\\
&\tilde n= \frac{(n_A+1)(n_B+1)(n_C+1)(n_A+n_B+n_C+1)}{(n_A+n_B+1)(n_A+n_C+1)(n_B+n_C+1)} \nn \ .
\end{align}

\begin{figure}[t]
\centering
\includegraphics[scale  = 1]{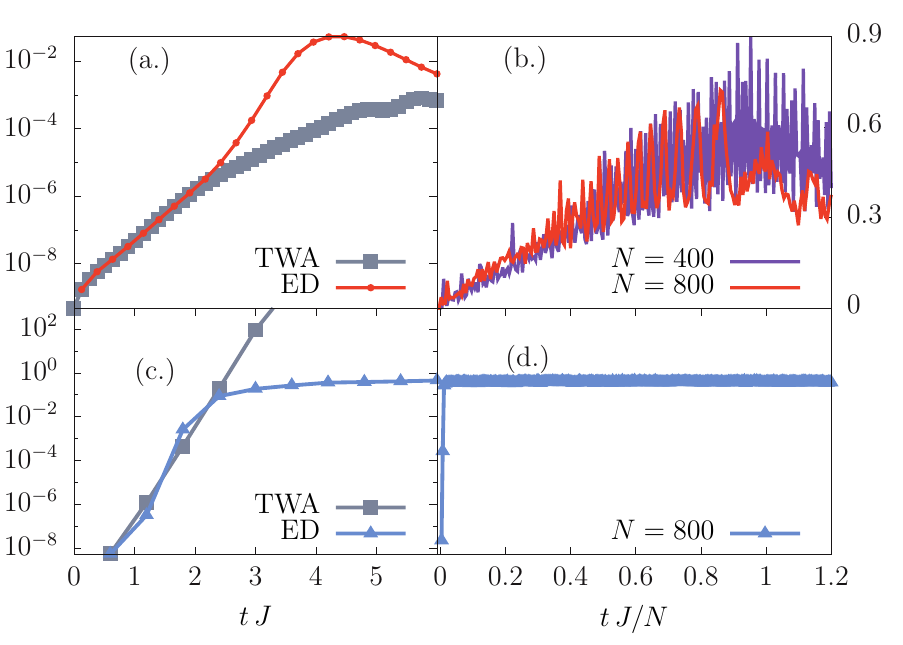}
\caption{Square commutator dynamics $c(t)$ for the LMG model after a quantum quench to the DPT point (a., b.) and for the kicked top with $K=20, \, h_f=2, \, 
\tau=0.6$ (c., d.). At early times (a., c.), they are both characterized by an exponential growth up to $t^c_{Ehr}$, see the right side with the 
log-scale on the $y$ axis.  This regime can be perfectly reproduced by the TWA. (b., d.)  At long times the 
behaviour is different depending on the quantum integrability properties. In (c.) the time is rescaled by $N$ in order to show that $c(t)$ grows up to a maximum at 
$t^r_*$. In (d.) In the Kicked top the $c(t)$ stays constant also at very long times. In (a., c.) we used $N=800$ and TWA obtained with $10^4$ samplings}
\label{fig:otoc_kicked_comparison}
\end{figure}

\subsection{Scrambling} 
\label{sec:scrambling}

 Scrambling, as measured by the square commutator, behaves in a profoundly different way from entanglement.  It is characterised 
by an initial semi-classical regime and a second quantum non-perturbative growth. 
Interestingly, this phenomenon is very evident in the regular regime 
(Fig.\ref{fig:otoc_class}), and it is much less clear in the chaotic one (Fig.\ref{fig:otoc_kicked_comparison}) already discussed at the beginning of the paper.

In the case of the  quench dynamics, for $h_f\gg h_c$ the square commutator is characterised by a first semiclassical quadratic growth  
$c(t)\propto {t^2}/N^3$ until $t^r_{Ehr}$. In this regime, semiclassical approximations describe very well the evolution of $c(t)$ and we chose to employ DTWA. To this end, we generalised the corresponding expression for the square commutator to the discrete phase space 
representation
\begin{equation}
	c(t) \sim \frac{\hbar^2}{N^4}  \overline{
    	\sum_{i, j, k, m} 
    	\left [\delta_{ij}^{xzy}(t) -  \delta_{ij}^{yzx}(t) \right ]\left [\delta_{km}^{xzy}(t) -  \delta_{km}^{yzx}(t) \right ]}
\label{eq:otoc_DTWA}
\end{equation}
\ni where $\delta_{ij}^{\alpha \beta \gamma}(t) =  \sigma_i^{\alpha}(0) \frac{\partial \sigma_j^{\beta}(t)}{\partial \sigma_i^{\gamma}(0)}$, with  
$\sigma^{x, y, z}_i$  the Weyl transform of the spin operators and the average $\overline {(\cdot)}$ is computed over the 
initial discrete Wigner distribution, see the supplementary material. At $t^r_{Ehr}$ the quantum regime starts, characterised by a polynomial growth $\sim (t/N)^4 $ up to a 
maximum $c(t^r_*) \sim 2$. At this time, the square commutator is independent of the system size.  Even the DTWA, that perfectly reproduces multipartite 
entanglement dynamics up to $t^r_{\text{rec}}$ (see Fig.\ref{fig:1}) is not able to reproduce the long time dynamics of the square commutator, 
see Fig.\ref{fig:otoc_class}. Indeed DTWA, despite keeping $N$ discrete trajectories, represents all operators as factorised on each site at any 
time~\cite{schachenmayer2015many}.  At times longer than $t_{\text{Ehr}}$, the operator expansion starts developing longer and longer strings and the 
square commutator re-sums all the correlations, until $t^r_*$, which corresponds to the time at which the string of length $N$ occurs. 
Quenches to $h_f\ll h_c$ are characterized by the same time-scales $t^r_{Ehr}$ and $t^r_*$ and the same semi-classical regime 
$\sim t^2/N^3$ up to $t^r_{Ehr}$. The result at long-times is qualitatively different: the quantum regime is $\sim t^{3}/N^4$ and $c(t^r_*)\sim 10^{-3}/N$ 
goes to zero at all times in the thermodynamic limit. This is a direct consequence of the existence of the dynamical transition, which is detected by scrambling \cite{heyl2018detecting}. Due to the presence of a macroscopic magnetization, the support of the operators has a constrained dynamics and it will not acquire a string of length $N$.

A special case is represented by the quench at  $h_f=h_c$; despite the integrability of the quantum system, we find that the square commutator grows exponentially in time up to $t^c_{Ehr}$ as $c(t)= e^{2t}/N^3$. This is due to the existence of the unstable trajectory in the classical dynamics. 
The exponent is twice  the eigenvalue of the instability matrix  of the separatrix trajectory $\lambda_{h_c}=2\sqrt{h_c(1-h_c)}$ 
for $h_c=1/2$.  This is valid in general for all the classical trajectories associated with DPT.  To our knowledge, it is the only example of an early time exponential growth in a many-body regular system. After $t^c_{Ehr}$, $c(t)$ keeps growing linearly in time up to the $t_*^r$ and then it goes back, 
see Fig.\ref{fig:otoc_kicked_comparison}.
Long-range interactions do not change drastically this finding. In the range $\a<1$, the early time dynamics is the same as that described before. 
The square commutator grows like a power law at small times, even for $\a>2$.

We conclude by considering the kicking, which induces a chaotic dynamics. As expected, $c(t)$ is initially dominated by the classical exponential 
growth, then, as $t\sim t^c_{Ehr}$, quantum interference effects appear and the square commutator saturates to a constant value~\cite{cotler2017out}, 
see Fig.\ref{fig:otoc_kicked_comparison} (lower panels).  In the quantum chaotic regime, the dynamics are reproduced by the semiclassical approximation, 
which predicts the initial time growth of the square-commutator. After $t^c_{Ehr}$, the TWA loses any physical meaning. The quantum $c(t)$ remains constant and finite in the thermodynamic limit, meaning that the operator's support is spread up to the longest string already from 
$t^c_{Ehr}$.  Notice that in this case, the exponent is different from the actual classical Lyapunov exponent, defined as  
$\lambda_{L}\equiv\overline{ \lim_{T\to \infty}\lim_{\delta Q_0\to 0} \log(|\delta Q(t)/\delta Q_0|)/T} $, that we compute following~\ocite{benettin1976kolmo}. 
As pointed out by~\ocite{rose2017classical}, this difference comes from a different ordering in evaluating the phase-space averages. \\

\section{Conclusions}
\label{sec:Conclusions} 
 In this work, we performed an analysis of the spreading of multipartite entanglement in long-range spin systems in comparison to that of scrambling. 
We show that quantum correlations build up and spread in different ways. While entanglement and the delocalization of the state's information are state properties, scrambling instead describes the growth of quantum correlations in operator's space.\\
After a quantum quench in the transverse field of the infinite range hamiltonian, entanglement is seen to saturate and become extensively multipartite $\sim \phi_Q\, N$ at $t_{\text{Ehr}}$. On the same timescale, the TMI $I_3$ saturates to a positive value $I_3>0$, showing that the information of the initial state is not delocalized across the system. In this case, entanglement dynamics is reproduced, up to very long times, by all the semiclassical approaches. 
On the other end, scrambling, as characterized by the square commutator of collective spin operators, increases semiclassically up to $t_{\text{Ehr}}$, continuing its growth even afterward up to the recurrence time. Particularly interesting is that, despite the dynamics being regular, the early-time exponential behaviour emerges when the dynamics occur at the critical point of the dynamical phase transition, see Fig.\ref{fig:otoc_kicked_comparison}. This point is associated with an unstable trajectory on the effective classical phase-space~\cite{sciolla2011dynamical} and the exponent of the square commutator is twice the eigenvalue of the instability matrix of the separatrix trajectory. Nonetheless, being the quantum dynamics regular, after $t_{\text{Ehr}}$, the square commutator keeps growing linearly in time up to the regular $t^*$, see Fig.\ref{fig:otoc_kicked_comparison}. The other quenches, as in Fig.\ref{fig:comparison}, are characterized by a first growth $\propto t^2/N^3$ up to $t_{\text{Ehr}}$ which is followed by a polynomial quantum
regime $\sim (t/N)^4 $ up to $t^*$. \\
In the presence of a periodic kicking for the chaotic evolution, entanglement saturates at $t_{\text{Ehr}}\sim \log N$ to values compatible with the infinite temperature state. As far as scrambling is concerned, we recover the exponential growth of the square commutators expected for chaotic systems up to $t_{\text{Ehr}}$, followed by the corresponding saturation induced by quantum interference effects, see Fig.\ref{fig:otoc_kicked_comparison}.\\
We conclude the summary of our results, by considering the case in which a quantum quench is performed with $\a\neq 0$. In this case, we employ TDVP and semi-classical analysis. Entanglement has the same asymptotic structure and dynamics within the interaction range $0\leq\a<1$: the QFI grows linearly in time up to a value $\sim N$, and the TMI increases logarithmically in time up to a constant value. For $1\leq\a<2$, the QFI and the TMI grow linearly in time and the entanglement structure of the asymptotic state is the same as for $\a<1$. Decreasing the range of interaction the situation changes: for $\a\geq2$ the state displays the typical dynamics and structure of short-range interacting systems 
$\sim \text{const}$; interestingly ${I_3}<0$ signaling that the information about the initial condition is spread throughout the degrees of freedom of the state (see Fig.\ref{Fig:bojan}). Coming to scrambling, we find that before $t_{\text{Ehr}}$ for $\a<1$, the square commutator grows as $\propto t^2$.

Despite entanglement can be efficiently reproduced up to very long times by our numerical tools, they all fail in predicting scrambling after $t_{\text{Ehr}}$. In our understanding, this follows from the fact that all methods approximate the support of the operator to remain factorized on the initial basis. Such approximations do not allow to reproduce the non-local behavior of scrambling at long-times.   
We would like to stress that the quantum regime of scrambling, which arises after $t_{\text{Ehr}}$ is not peculiar only of our models or of regular dynamics.  In fact, a power law in the quantum regime has
been found also in chaotic systems~\cite{bagrets2016syk,bagrets2017power}. The long-time behavior of the square commutator shows the presence of purely quantum correlations that build up in the operator space. In this respect, it would be interesting to explore new quantum information protocols that encode information in the operator itself.

\acknowledgments

 We acknowledge useful discussions with J. Goold, A. Lerose and A. Polkovnikov. We acknowledge  support
from EU through project QUIC under grant agreement 641122, the National Research Foundation of Singapore
(CRP - QSYNC). B. Z. is supported by the Advanced grant of the European Research Council (ERC), No. 694544 – OMNES.


\widetext

\begin{center}
\textbf{\Large Supplementary Material: \\ \medskip
\Large Scrambling and entanglement spreading in long-range spin chains }
\end{center}
\setcounter{equation}{0}
\setcounter{figure}{0}
\setcounter{table}{0}

\setcounter{page}{1}
\makeatletter
\renewcommand{\theequation}{S\arabic{equation}}
\renewcommand{\thefigure}{S\arabic{figure}}
\renewcommand{\bibnumfmt}[1]{[S#1]}
\renewcommand{\citenumfont}[1]{S#1}

In Section 1, a brief recap on the semi-classical numerical approximate methods used to reproduce the square-commutator dynamics. In Section 3, we report some more plots of the entanglement dynamics.\\

\twocolumngrid
\section{Details on truncated approximations}
\label{app:trunca_appro}

\subsection{Truncated Wigner Approximation (TWA)}
Wigner formalism is based on a mapping between the Hilbert Space of a quantum system and its corresponding phase space, known as the Wigner-Weyl transform. This is achieved through the so-called \emph{phase-point operator} $\hat A(\bold q, \bold p)$, where $\{\bold q, \bold p\}$ are the classical phase-space variables \cite{wootters1987discrete}. Operators $\hat O$ are mapped to functions on phase-space: $O^w(\bold q, \bold p) = \tr[\hat O\, \hat A(\bold q, \bold p)] $, known as the Weyl symbols. The Weyl symbol of the density matrix $\hat \rho$ is called \emph{Wigner function} $W(\bold q, \bold p) =  \tr[\hat \rho\, \hat A(\bold q, \bold p)]$. This inherits the density matrix hermiticity and normalization, being a quasi-probability distribution, in general non-positive. Within this frame, it is possible to compute time-dependent expectation values as weighted averages over phase space of the Weyl symbols as
\begin{align}
\langle \hat O(t)\rangle &
    = \tr[ \hat \rho_0 \, \hat O(t) ] 
    = \int d\bold q_0 \, d\bold p_0 \, W(\bold q_0, \bold p_0)
        \, O^w(\bold q(t), \bold p(t) )  
\end{align}
where the weigh is given by the initial Wigner function. When quantum fluctuations can be neglected \cite{SI_polkovnikov2010phase}, the Weyl symbol can be evaluated over the classical trajectories as 
\begin{equation}
\langle \hat O(t)\rangle 
     \simeq \int d\bold q_0 \, d\bold p_0 \, W(\bold q_0, \bold p_0)
        \,  O^w(\bold q_{cl}(t), \bold p_{cl}(t) ) \ ,
\end{equation}
 This approximation is known as the Truncated Wigner Approximation. When the $W(\bold q_0, \bold p_0)$ is positive, it can be interpreted as a probability distribution and from a numericla point of view it is possible to consider a Montecarlo sampling 
\begin{equation} \label{eq:TWA}
\langle \hat O(t)\rangle_{TWA}\simeq 
    \frac 1{N_{random}}
    \sum_{i=1}^{N_{random}}
    O^w(\bold q^i_{cl}(t), \bold p^i_{cl}(t) )
\ , 
\end{equation}
where $\bold q^i_{cl}(t)$ and $\bold p^i_{cl}(t)$ are the classical trajectories corresponding to the $i-$th initial condition randomly distributed according to the initial Wigner function. \\ 

\begin{figure}[t!]
\centering
\includegraphics[scale  = 1]{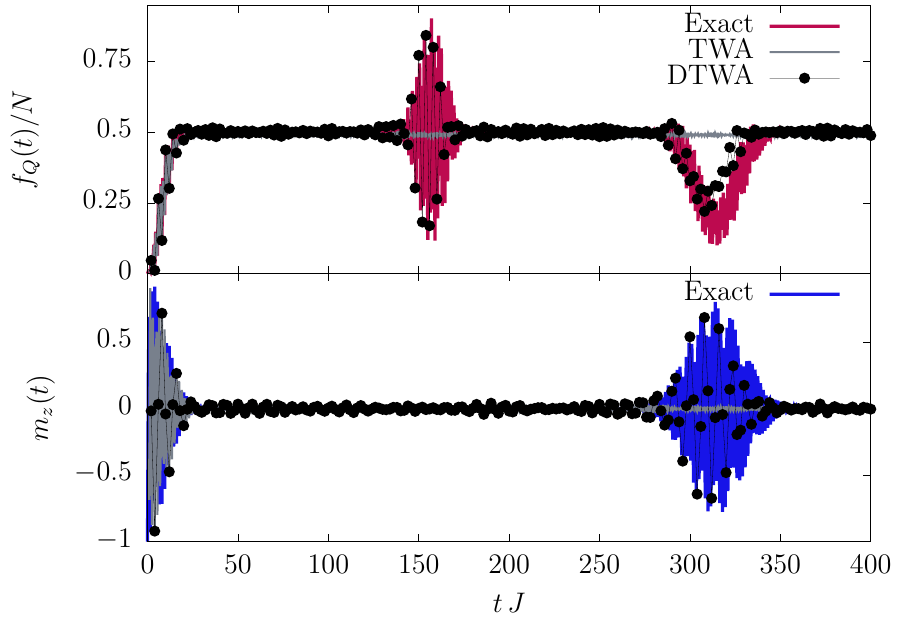}
\caption{Comparison between the different semiclassical approaches to the dynamics of local observables (bottom) and the multipartite entanglement (top). For $N=100$ and $h_f=2$. TWA and DTWA obtained with $N_{random}=5\, 10^3$ samplings. In contrast with the other methods, DTWA is able to reproduce $t^r_{\text{rec}}$ and beyond.  }
\label{fig:enta_satu}
\end{figure}

In our semiclassical model, the TWA consists in expressing the observables in terms of the magnetization's components $\bold m$, in evolving them according to the classical equation of motion Eq.(6) and then in averaging over different initial conditions sampled according to $W(\bold m_0)$.  
The square commutator is computed as in Eq.(9), where the average is taken over the Wigner function of the initial state. The initial state $\ket{\psi_0}=\ket{\up\, \up\,\dots \up\,}$ gives a gaussian for the transverse components, with variance $\sigma=\frac 1{\sqrt S}$ and the component along $z$ is fixed by the conservation of the total spin: $\bold m\cdot \bold m=1$: the Wigner function reads $W(\bold m_0)=\frac 1{\pi S} \,e^{-(m^2_{0\,x}+m_{0\,y}^2)\,S}\, \delta(m_{0\,z}-\sqrt{1-m^2_{0\,x}+m_{0\,y}^2})$, see \cite{polkovnikov2010phase}. This approximation treats the quantum degrees of freedom collectively. For this reason it reproduces the observable's dynamics before the Ehrenfest time, but is not able to capture long-time-dynamics and the revivals neither of the magnetization and entanglement, see Fig.\ref{fig:enta_satu}, nor the long-time behavior of the square-commutator, see Fig.3.

\medskip
\subparagraph{Discrete Truncated Wigner Approximation (DTWA)} 
The previous mapping can be generalized to systems with $n$ discrete degrees of freedom \cite{wootters1987discrete}, where the continuous phase space is replaced with a discrete one. Consider an example for $n=2$ one spin-$1/2$. It can be described in a discrete phase space made of $n^2=4$ points, each of them associated with coordinates $\alpha = (a_1, a_2)\in\{(0,0), (0,1), (1, 0), (1,1)\} $. Notice that this can be generalized to $n>2$, with $n$ prime. For a beautiful and clear survey of the topic see \cite{wootters1987discrete}. Let $\hat A_{\alpha}$ be the discrete phase point operator associated to the point $\alpha$. As in the continuous case, each operator implements the mapping between the Hilbert space of the quantum system and this discrete phase space. One can show that, by choosing $\hat A_{\alpha}$ in an appropriate way, all the properties of the phase space operator that hold in the continuum case (as the normalization or the orthonormality) still hold. Such a choice is achieved by
\begin{align}
\hat A_{\alpha} & = \frac 12 \left [
    (-1)^{a_1} \hat \sigma_z 
    + (-1)^{a_2} \hat \sigma_x 
    + (-1)^{a_1 + a_2} \hat \sigma_y
    + \mathbb I
    \right ] \nn \\
    & = \frac 12 \left (
    \mathbb I + \bold r_{\alpha} \cdot \pmb {\hat \sigma}
    \right ) \ ,
\end{align}
with $\bold r_{\alpha} =((-1)^{a_2}, (-1)^{a_1+a_2}, (-1)^{a_1})$ and $\pmb {\hat \sigma}=(\hat \sigma_x, \hat \sigma_y, \hat \sigma_z)$ are the Pauli matrices. Then one can construct the Weyl symbol of a generic operator as before: $O^w_{\alpha} = \tr(\hat O \hat A_{\a})/2$. The discrete Wigner function $w_{\a}=\tr(\hat \rho \hat A_{\a})/2$ can be pictured as a $2\times 2$ matrix, see Eq.(\ref{eq:ini_cond}), and it gives the probability that a state is in the point $\alpha$. Moreover, following the construction of $\hat A_\a$ from \cite{wootters1987discrete}, the sum over the horizontal lines of $w_\a$ gives the probabilities for the $z$ component of the spin, the sum over the vertical lines gives the probabilities for the $x$ components and the sum on the diagonal ones gives the $y$ component probabilities.  As example, consider a spin pointing along the $z-$direction: $\hat \rho = \ket {\up\, }\bra{\, \up}$. Its description in the discrete phase space will be
\begin{equation} \label{eq:ini_cond}
w_{\a} =
\begin{tabular}{| l |c | } 
       \hline
        $w_{00}$ &  $w_{01}$ \\
        \hline  
        $w_{10}$ &  $w_{11}$ \\  
    \hline
\end{tabular}
= 
\begin{tabular}{| l |c | } 
       \hline
        $\, 1/2 \, $ &  $\,  1/2\, $ \\
        \hline  
        $\,\,\,\,0\, $ &  $\,\, 0\, $ \\  
    \hline
\end{tabular}
 \ ,
\end{equation}
therefore the state will point along $z$ with probability one, while the probability of being along $+x$ or $-x$ (or along $+y$ or $-y$) will be $1/2$.

If now we consider a composite system made of $N$ spins $1/2$, the previous description can be immediately generalized. In this case, the space is represented by $4^N$ configurations: $\pmb{ \a} =\{\a_1, \a_2, \dots, \a_N\}$. The phase-point operator is given by the tensor product of the single-sites one, as:
\begin{equation}
\hat A_{\pmb{ \a}} = \hat A_{\a_1} \otimes \hat A_{\a_2} \otimes \dots \otimes \hat A_{\a_N} \ .
\end{equation}
The expectation values of operators are written as 
\begin{equation}
\label{eq:operator_decompose}
\langle \hat O\rangle = \sum_{\pmb \a} w_{\pmb \a}\, O^w_{\pmb {\a}} \ ,
\end{equation}
with the discrete Wigner-function $w_{\pmb \a}\equiv \frac 12 \Tr[\hat A_{\pmb \a} \hat \rho]$, where now $\hat \rho$ is the many-body density matrix and the Weyl transform of the operator: $O_{\pmb \a}^w \equiv \Tr[\hat O\, \hat A_{\pmb {\a}}]$. Notice that for pure states $\hat \rho$ is a sum of projectors, hence $w_{\pmb \a}$ factorizes too. In particular for initially separable states, $w_{\pmb \a}$ factorizes in the product of $N$ independent spin-$1/2$ $w_{\a_j}$ with $j=1, \dots, N$. Indeed, the initial state $\ket{\psi_0}=\ket{\up\, \up\,\dots \up\,}$ will be mapped into $w_{\pmb \a}(0) = \prod_{i=1}^N \, w_{\a_i}^{[i]}$ with $w_{\a_i}^{[i]}$ equal to Eq.(\ref{eq:ini_cond}) $\forall i$. Furthermore notice that $\hat O=\hat o_j$ is a single site operator, then a useful simplification occurs:  $(o_j)^w_{\pmb \a} =(o_j)^w_{\a_j}$.

The Discrete Truncated Wigner Approximation is equivalent to the TWA, but on this discrete phase space representation \cite{schachenmayer2015many}. Hence, starting from Eq.(\ref{eq:operator_decompose}), one evolves classically the Weyl symbol of the operator and averages over the initial Wigner function as
\begin{align}\label{eq:dtwa}
    \langle \hat O(t)\rangle = \sum_{\pmb {\a}} w_{\pmb {\a}}(0)\, O^w_{\pmb {\a}}(t) 
    \simeq    
    \sum_{\pmb {\a}} w_{\pmb {\a}}(0)\, O^{w, cl}_{\pmb {\a}}(t)\\ \nn
    \simeq
    \frac 1{N_{random}} \sum_{m}^{N_{random}}\,  O^{w, cl}_{\pmb {\a}_m}(t) \ ,
\end{align}

where we extract $N_{random}$ initial spin configurations according to the initial Wigner transform. 
The usual TWA integrates classical trajectories on phase space and then averages over initial conditions, by contrast, DTWA discretizes the initial conditions and then evolves them classically. As for the TWA, we compute local observables, the QFI and the square commutator (see next paragraph) and we compare it with the results we get with the TWA, Fig.\ref{fig:enta_satu}. \\
In order to compute time-ordered correlators, one has to compute the Weyl symbol of operators like $\hat O = \hat m^{z} = \frac 1{N}\, \sum_{j=1}^N\, \hat \sigma_i^z$ and his powers. The Weyl's symbol of $\hat m_z$ is immediate 
\begin{align}
    m^{z,\, w}_{\pmb {\a}}(t) & = \Tr\left [  \hat m^z(t)\, \hat A_{\pmb {\a}}  \right ]
    = 
    \frac 1 N \sum_{j=1}^N\, \Tr\left [\hat \sigma_j^z(t) \, \hat A_{\pmb {\a}}  \right ] 
    \nn \\ & \simeq
    \frac 1 N \sum_{j=1}^N\, \Tr\left [\hat \sigma_j^z(t) \, \hat A_{\a_j}  \right ] 
    =
    \frac 1N  \sum_{j=1}^N \, \sigma^{z, \, \a_j}_j(t) \ ,
\end{align}
defining $ \sigma^z_j = \Tr\left [\hat \sigma_i^z \, \hat A_{\a_j}  \right ]$ as the Weyl symbol of the single site magnetization, which evolves with the classical equation of motions. The dependence on $\a$ remains in the choice of the initial conditions. Analogously, the Weyl transform of $\hat O = \hat m^{z\, 2}= \frac 1{N^2}\, \sum_{ij}\hat \sigma_i^z\hat \sigma_j^z$
\begin{align}
    m^{z\, 2,\, w}_{\pmb {\a}} & = \Tr\left [  \hat m^{2\, z}\, \hat A_{\pmb {\a}}  \right ]
   = 
    \frac 1 {N^2} \sum_{i,j=1}^N\, \Tr\left [\hat \sigma_i^z \, \hat \sigma_j^z \,\hat A_{\pmb {\a}}  \right ] \nn \\
    & =
    \frac 1 {N^2} \sum_{i,j=1}^N\, \Tr\left [\hat \sigma_i^z \, \hat \sigma_j^z \,\hat A_{\a_i} \hat A_{\a_j} \right ] 
    = \frac 1{N^2}  \sum_{i,j=1}^N \, \sigma^{z,\,\a_j}_j \, \sigma^{z,\,\a_i}_i \ .
\end{align}
Notice that an important approximation has been made: we took the support of $\hat \sigma_i^z(t)$ to be localized on the site $j$: this allows to write $\Tr\left [\hat \sigma_j^z(t) \, \hat A_{\pmb {\a}}\right] \simeq \Tr\left [\hat \sigma_j^z(t) \, \hat A_{\a_j}  \right ]$. This approximation is exactly the same as shifting the time dependence on the phase-space operator and then to approximate it as factorized at every time $t$, as done in \cite{schachenmayer2015many}
\begin{equation}
\hat A_{\pmb \a}(t) \simeq 
    \hat A_{\a_1}(t) 
    \otimes \hat A_{\a_2} (t)
    \otimes \dots 
    \otimes \hat A_{\a_N}(t) \ .
\end{equation}
As known, the DTWA works incredibly well for local observables as $\langle \hat m_z(t)\rangle$ and the QFI \cite{schachenmayer2015many}. It is also able to reproduce the long-time behavior and the recurrences, see Fig.\ref{fig:enta_satu}. 
We verified the long-time validity also for correlators at different times, like $\langle \hat m_x(0) \hat m_z(t)\rangle$.

\section{Entanglement dynamics plots}
\label{app:enta}
\begin{figure}[h]
\centering
\includegraphics[scale  = 1]{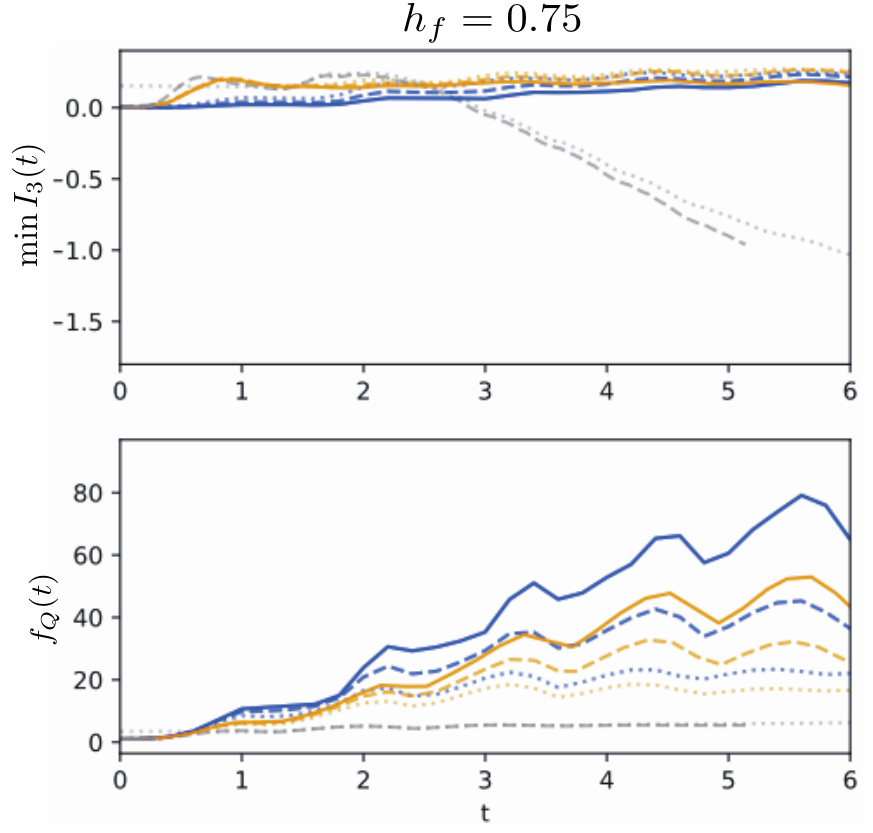}
\caption{Scaling of the QFI and the tripartite mutual information with the system size for long-range hamiltonians for $\a \neq 0$. We observe that whenever the $f_Q(t)$ increases with the system size the minimum of the tripartite mutual information becomes independent of the system size and vice versa. The colors correspond to different interaction ranges: $\alpha=2.5$ (gray) $\alpha=1.5$ (orange) and $\alpha=0.5$ (blue). The brightness and line style correspond to different system sizes $n=50, 100, 200$ from bright to dark (or dotted to full lines). All data is converged with the bond dimension $D=256$, except the data for $\alpha=2.5$ where bond dimension $512$ had to be used.}
\label{fig:boyan}
\end{figure}

\begin{figure}[h]
\fontsize{12}{10}\selectfont
\centering
\includegraphics[scale  = 1]{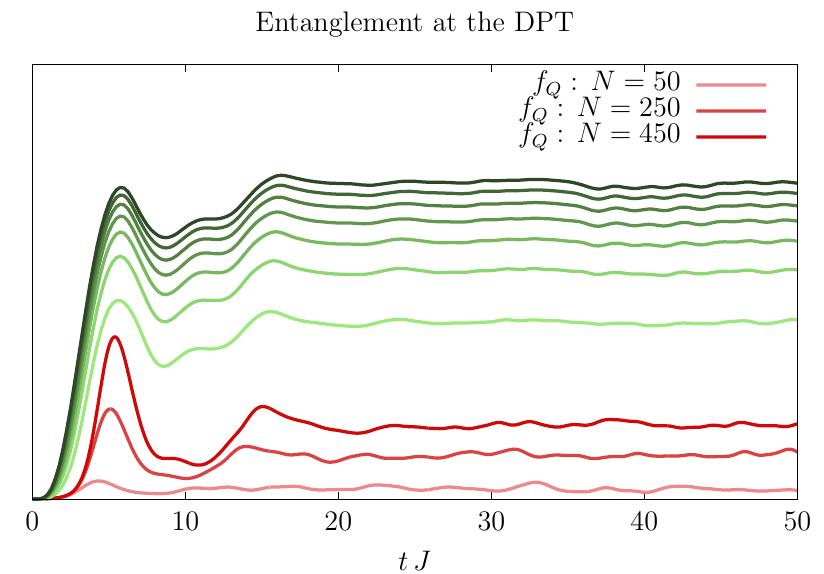}
\includegraphics[scale  = 1]{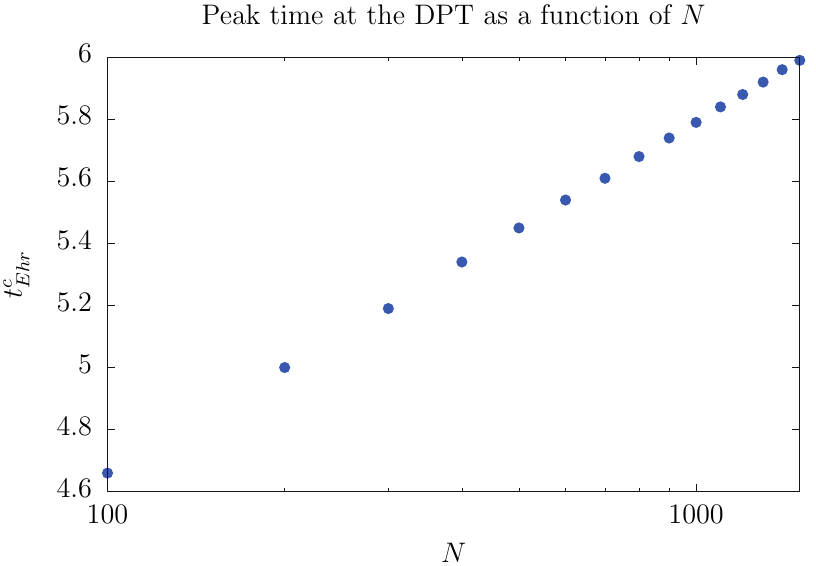}
\caption{Entanglement dynamics at the DPT, $h_0=0$ and $h_c$. (Left panel) Bipartite and multipartite entanglement dynamics at the DPT in two different scales. \textbf{Green}: bipartite entanglement $S_L(t)$ for $N=500$ with $L=10, \dots, 70$. It is evident that $t^c_{Ehr}$ shifts with $N$, but does not depend on $L$. \textbf{Red}: multipartite entanglement $f_Q(t)$ with increasing values of $N$. The peak scales linearly with time and it reaches a value $f_Q(t^c_{Ehr})= 0.65 \, N$.  (Right panel) Scaling of $t^c_{Ehr}$ with the system size $N$. Note the log-sale on the axis $N$, meaning that the time for which we obtain a maximum scales logarithmically  with $N$.}
\label{fig:enta_dpt}
\end{figure}

\begin{figure}[H]
\fontsize{12}{10}\selectfont
\centering
\includegraphics[scale  = 1]{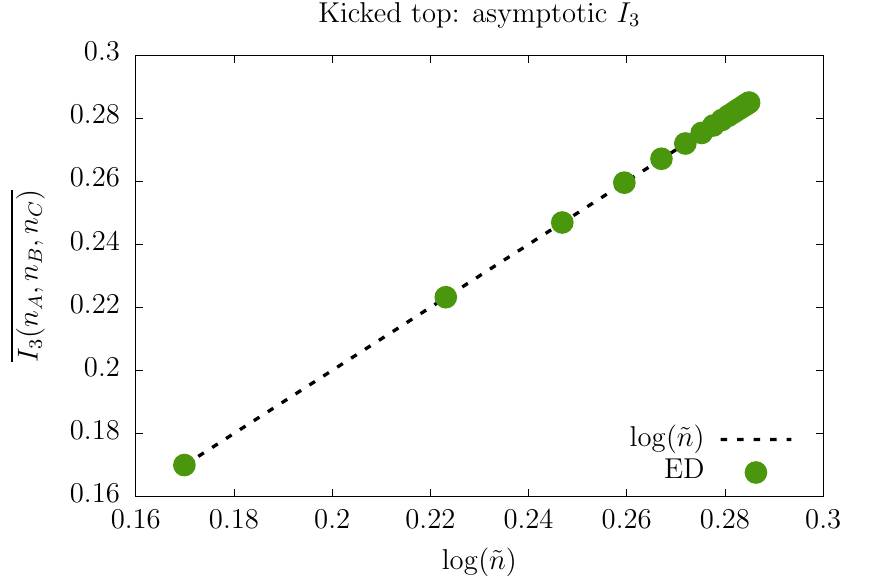}
\includegraphics[scale  = 1]{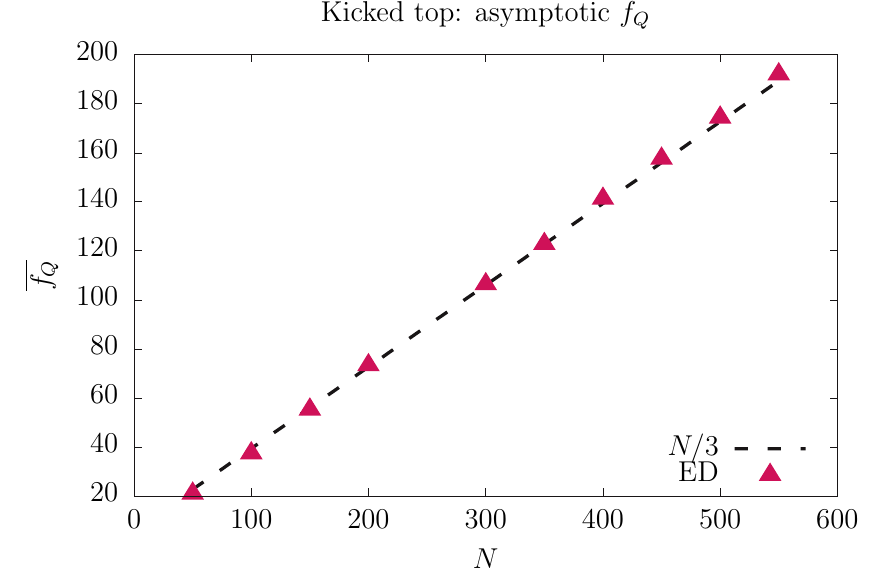}
\caption{Scaling of the entanglement saturation in the chaotic regime. It is well known that the asymptotic state in the chaotic regime ($K=20$) can be described by a state at infinite temperature: a uniform superposition of all the states in the allowed Hilbert space \cite{russomanno2015thermalization}. This reflects also on the entanglement properties of the asymptotic state. In fact, for every initial states and for all $h$ and $K$ in the chaotic region, after few kicks the entanglement quantities saturates to their ergodic values.
}
\label{fig:enta_chaos}
\end{figure}

In this section of the supplementary, we report some more plots of the entanglement dynamics. The comparison between QFI and TMI is considered in the case of a quantum quench to the dynamical quantum phase transition in Fig.\ref{fig:enta_dpt}, with long-range hamiltonians at $\a\neq 0$ in Fig.\ref{fig:boyan} and for a chaotic kicked dynamics in Fig.\ref{fig:enta_chaos}.

\end{document}